\documentclass[sigconf, natbib=false, nonacm, authorversion]{acmart}

\usepackage{amsmath,amsfonts}
\usepackage{mathtools}
\DeclarePairedDelimiter{\set}{\{}\}
\usepackage{algpseudocode}
\usepackage{algorithm}
\usepackage{graphicx}
\usepackage{textcomp}
\usepackage{xcolor}
\usepackage{multirow}
\usepackage{makecell}
\usepackage[inline]{enumitem}

\usepackage[datamodel=acmdatamodel, style=acmnumeric, doi=false, url=false, date=year, isbn=false, eprint=true]{biblatex}
\AtNextBibliography{\small}
\AtEveryBibitem{\clearfield{pages}}
\AtEveryBibitem{\clearfield{pagetotal}}
\AtEveryBibitem{\clearfield{volume}}
\AtEveryBibitem{\clearfield{number}}
\AtEveryBibitem{\clearlist{language}}
\acmConference[Middleware26]{27th ACM/IFIP International Middleware Conference}{November 30--December 04,
  2026}{Tarragona, Spain}

\begin{document}

\title{BENNS: A Surrogate Model for Hybrid Online-Offline Evolution of SFC Embedding}

\author{Theviyanthan Krishnamohan}
\email{theviyanthan.krishnamohan@glasgow.ac.uk}
\orcid{0000-0003-0040-1130}
\author{Lauritz Thamsen}
\email{lauritz.thamsen@glasgow.ac.uk}
\orcid{0000-0003-3755-1503}
\author{Paul Harvey}
\email{paul.harvey@glasgow.ac.uk}
\orcid{0000-0003-1243-938X}
\affiliation{%
  \institution{University of Glasgow}
  \city{Glasgow}
  \country{United Kingdom}
}

\renewcommand{\shortauthors}{Krishnamohan et al.}

\begin{abstract}

Service Function Chains (SFCs) enable programmatic control of the functions and services in a computer network. By leveraging Software Defined Networking to control the links between virtualised network functions, SFCs provide a scalable approach to dealing with the increased pressures on network operation and management. However, embedding SFCs onto the underlying physical network and compute infrastructure is an NP-hard problem. Genetic Algorithms (GAs) have been used to address this issue, but they require significant time to evaluate solution quality (fitness) \textit{online}, with most existing approaches instead adopting \textit{offline} simulations or analytical evaluations.

To enable online use of GAs in solving the SFC embedding problem, we introduce a hybrid online-offline approach to efficiently evaluate the fitness of generated solutions. At the core of this is BENNS: a surrogate model that approximates fitness and is agnostic to topology, traffic, and SFC-embedding. We evaluate our approach in a static environment across five experiments, varying available resources and traffic loads, and in a dynamic network environment. Our results demonstrate that our approach is capable of exploring thousands of potential configurations and generating deployable solutions in 19.1 minutes on average, compared to online-only approaches, which take 17.8 hours on average to explore ten solutions in our experiments and do not converge on an optimal solution.
\end{abstract}




\maketitle

\section{Introduction}\label{introduction}
Computer networks are essential for our economy, education, and social well-being~\cite{Pandav2022LeveragingCare., annurev:/content/journals/10.1146/annurev-financial-101620-063859, 6727567}, making it important that they operate reliably, without interruption, and at optimum quality. The pressures of increased user numbers and changing operational patterns make achieving this optimal operation challenging. Autonomous operation of networks is seen as the holy grail in achieving this uninterrupted operation in a scalable and resource-efficient manner~\cite{10255468, Imai2020TowardsNetwork, Yannan2020AutonomousBusiness}. 

Service Function Chaining (SFC)~\cite{Halpern2015} is essential for bringing autonomous networks closer to reality. SFC uses Network Function Virtualisation (NFV)~\cite{NFV2014} and Software Defined Networking (SDN)~\cite{2013Software-DefinedIt} to enable programmatic control of the configuration of a network and its services. NFV enables network functions, such as a firewall, to be virtually deployed in any server. These Virtual Network Functions (VNFs) can then be linked together using SDN, forming SFCs. 

However, this programmatic flexibility introduces the challenges of how to embed SFCs upon the physical network and compute infrastructure. This involves optimally ordering VNFs in an SFC (\textit{chain composition}), deploying VNFs in servers (\textit{VNF embedding}), and routing traffic among these VNFs (\textit{link embedding})~\cite{GilHerrera2016}. We refer to these optimisation problems collectively as the Optimal SFC Embedding (OSE) problem in the remainder of this paper. The OSE problem is NP-hard.

Genetic Algorithms (GAs) are metaheuristic algorithms that have been shown to be effective for NP-hard problems~\cite{Acampora2025SolvingAlgorithms}. GAs involve evolving thousands of candidate solutions, using evaluations of their \textit{fitness} to guide successive evolutions. The evaluation can be done \textit{online}, where solutions are evaluated by experimentation, or \textit{offline}, where fitness is evaluated numerically or using simulators. Online evaluation is more accurate and captures the complexity of modern networks~\cite{10255468}, but is time-consuming~\cite{Mori2000GeneticSurvey}. Offline evaluation trades accuracy for speed~\cite{OpenRASE, Mori2000GeneticSurvey}. We call evolving solutions using online evaluation \emph{online evolution}, and offline evaluation \emph{offline evolution}. 

Of the 19 studies found that apply GAs to the OSE problem, one performs online evolution~\cite{OpenRASE}, while 18 perform offline evolution~\cite{Khoshkholghi2019,Carpio2017,10.1007/978-3-031-33743-7_39,Toumi2022,Li2019,Gamal2019,Gamal2019a,Qu2016,Khoshkholghi2020,Kim2016,Rankothge2017,Ruiz2020,Cao2017,Tavakoli-Someh2019,Ma2017,Fulber-Garcia2024,Yuan2018,Fulber-Garcia2023}, of which only two have a final partial verification step using an emulator. The online evolution established the higher fidelity representation of the network compared to offline evaluation, but at the cost of significant computational time. Offline approaches use user-specified resource demands of SFCs, as opposed to benchmarked data, limiting representational accuracy.

This paper presents a hybrid online-offline approach to evaluate the fitness of GA-based solutions to the OSE problem, addressing the time/accuracy trade-off. We propose a novel benchmarking and Neural Network(NN)-based \textit{surrogate model} (Section~\ref{sec:benns}) named \textbf{Be}nchmarking and \textbf{N}eural \textbf{N}etwork-based \textbf{S}urrogate (BENNS) that \textit{approximates} SFC performance on a network topology combined with online evolution using OpenRASE~\cite{OpenRASE}, an existing SFC emulator. Our approach is designed to be independent of the network topology, traffic and SFC embedding, and simple enough to run on CPUs, ensuring faster approximation. An SFC embedding consists of all the SFCs embedded in a network. To the best of our knowledge, we are the first to use a trained surrogate model for offline evaluation and a hybrid online-offline evolution approach to solve the OSE problem. 

We experimentally evaluate both the assumptions behind our model,s and our approach in a static environment across five different network scenarios and a dynamic environment, demonstrating that our approach is capable of exploring tens of thousands of possible configurations to produce an optimal solution faster, compared to online evolution, which was slower to explore ten solutions and did not produce any optimal solution.

\section{Related Work} \label{related}
To place our work in context, we analysed the state of the art to identify fitness evaluation approaches taken by GA-based solutions to the OSE problem. We also analysed non-GA solutions to OSE to identify evaluation mechanisms that may have a potential use in GAs. Finally, we also studied the use of GAs in other application domains to identify evaluation mechanisms that may be useful in the OSE domain. 

\subsection{GA-based Solutions to OSE}
Of the 19 studies found that explore the use of GAs to the OSE problem~\cite{Khoshkholghi2019,Carpio2017,10.1007/978-3-031-33743-7_39,Toumi2022,Li2019,Gamal2019,Gamal2019a,Qu2016,Khoshkholghi2020,Kim2016,Rankothge2017,Ruiz2020,Cao2017,Tavakoli-Someh2019,Ma2017,Fulber-Garcia2024,Yuan2018,Fulber-Garcia2023,OpenRASE}, only one used an online approach~\cite{OpenRASE}, while none used a hybrid approach. Several studies used tools to partially \textit{verify} the final solutions generated from offline evolution: Ruiz et al.~\cite{Ruiz2020} and Kim et al.~\cite{Kim2016} used the OMNET++ simulator, Tavakoli-Someh et al.~\cite{Tavakoli-Someh2019} used the CloudSim simulator, and Ma et al.~\cite{Ma2017} used a Java-based discrete event simulator to verify the solution produced by their GA. Simulators model reality and have less accuracy compared to online approaches~\cite{Fahmy2023SimulatorsWSNs}. So, we consider these as offline approaches. Khoshkholghi et al.~\cite{Khoshkholghi2020} and Rankothge et al.~\cite{Rankothge2017} used Mininet, an emulator which has higher accuracy~\cite{Fahmy2023SimulatorsWSNs}, to verify their solutions, but did not test their solution completely on the emulator. Rankothge et al.\ used Mininet to emulate the network topology only, while using a simulation prototype of their Network Function Centre to simulate network functions~\cite{Rankothge2017}. Similarly, Khoshkholghi et al.\ used Mininet to verify the performance of network links while relying on numerical analysis to evaluate the performance of network functions~\cite{Khoshkholghi2020}. The offline approaches in the literature did not validate their solutions on a real network or an emulator completely. Krishnamohan et al.~\cite{OpenRASE} performed online evolution using OpenRASE and showed it to be of higher accuracy than offline evolution, but the online evolution took orders of magnitude longer compared to offline approaches. Consequently, offline evolutions in the literature are significantly faster, whereas online evolution can yield considerably higher accuracy. However, this trade-off has not been exploited by any hybrid approach so far.

\subsection{Non-GA Solutions to OSE}
We analysed non-GA solutions to OSE to identify evaluation mechanisms that may be relevant to GAs. Even though other algorithms may not necessarily include fitness evaluation as a part of the algorithm, the mechanisms used to verify the solutions produced by these algorithms may be suitable for fitness evaluation in GAs. We found offline approaches in the form of simulators such as ALEVIN, OMNET++, and CloudSim~\cite{OpenRASE}, and online approaches in the form of emulators such as Mininet and OpenRASE, virtualisation tools such as Kubernetes, OpenStack, and OpenNetVM, and testbeds such as CREATE-NET, 5GInFIRE, and a Raspberry Pi-based platform~\cite{OpenRASE}. Since emulators are closer to reality compared to simulators~\cite{Fahmy2023SimulatorsWSNs}, they are considered to be online~\cite{OpenRASE}. Among online approaches, OpenRASE has been built for OSE evaluation and is scalable. The offline approaches, which are simulations, require users to specify the resource demands of SFCs and the numerical function to compute SFC performance. For instance, ALEVIN requires the CPU and bandwidth demand of SFCs to be user-defined, and performance has to be numerically evaluated~\cite{OpenRASE}. We also found numerical evaluations that quantify SFC performance using mathematical formulas based on assumptions about factors influencing SFC performance without validating the assumptions~\cite{Harutyunyan2022, Liu2021a, Gamal2019a, Fu2020, Tahmasebi2018, Liu2024a, Khoshkholghi2020}. Additionally, Bunyakitanon et al.~\cite{Bunyakitanon2020} trained a VNF performance prediction model of a video transcoder using data from a real network, which was used to optimise the OSE problem. 

\subsection{Offline Evolution in Other Applications}
Offline evolution in other application domains takes the form of \textit{surrogate models}~\cite{Miller2024OptimizingFidelity} that are used to quickly approximate the fitness of candidate solutions, trading accuracy for time. Popular mechanisms are polynomial models, Kriging models, and NNs~\cite{jin2002fitness}. In addition, fitness inheritance, clustering techniques, and Support Vector Machines (SVM) have also been used~\cite{Shi2010AAlgorithms}. The performance of Kriging and SVM depends on the kernel used, and selecting the right kernel is challenging~\cite{Oyetunde2022NavigatingSolutions}, while polynomial models require polynomial relationships. Clustering techniques and fitness inheritance involve multiple online evaluations, which increases convergence time.

NNs have become a popular surrogate model in recent times. Pan et al.~\cite{Pan2019AOptimization} used an NN to predict the dominance relationship between individuals and their references in a many-objective evolutionary algorithm. An NN was used to build a surrogate model for a multi-objective composite shell structure optimisation problem by Miller et al.~\cite{Miller2024OptimizingFidelity}. Dushatskiy et al.\ used a Convolutional Neural Network as a surrogate model in a Gene-pool Optimal Mixing Evolutionary Algorithm~\cite{Dushatskiy2019ConvolutionalGOMEA}. NNs have also been used as surrogates in optimal aerodynamic designs~\cite{Sun2019ADesign}. The use of an NN as a surrogate model in optimising multi-generation energy systems was explored by Ghafariasl et al.~\cite{Ghafariasl2024NeuralSystem}. Sreekanth et al.~\cite{SREEKANTH2010245} used an NN-based surrogate model to optimise multi-objective saltwater intrusion management in coastal aquifers. An NN model was also used in the evolutionary optimisation of catalytic materials~\cite{Holena2010NeuralAlgorithms}.

\section{OSE Problem Definition}\label{prob_def}
In this work, we consider a data centre environment of a cloud service provider that provisions SFCs to clients, a popular SFC deployment context~\cite{Xie2021, Jang2017, Xu2021, Wang2020a}. Given that OpenRASE supports only HTTP traffic, we consider only web-based traffic. We consider a set of SFC Requests (SFCRs) to be embedded in the data centre such that the \textit{acceptance ratio} is \textit{maximised} and the \textit{average traffic latency} is \textit{minimised}. These are the optimisation objectives of the OSE problem and become the \textit{fitness scores} of individuals in the context of GAs. Acceptance ratio is the number of SFCs that can be deployed divided by the number of SFCRs. Traffic latency is the round-trip time traffic takes to traverse an SFC. We average the traffic latency across all SFCs during the entire duration of traffic to find the average traffic latency of an SFC embedding. These are conflicting objectives. The more we embed SFCs, the higher the competition for resources, which makes minimising average traffic latency challenging, making this a Pareto optimisation problem. Acceptance ratio can be calculated mathematically; however, traffic latency has to be measured. Consequently, the surrogate model has to approximate \textit{traffic latency only}.  

\section{Hybrid Online-Offline Evolution}
To address the tradeoff between accuracy and time in evaluating GA-generated solutions to the OSE problem, we now describe the overall architecture to enable a hybrid online-offline evolutionary approach (Fig.~\ref{fig:hybrid}) blending the efficiency and sufficient accuracy of the BENNS surrogate model with the slower but higher accuracy of OpenRASE, the emulator. 

\subsection{Approach}
\begin{figure}[t]
    \centering
    \Description{A diagrammatic representation of hybrid evolution using BENNS and OpenRASE}
    \includegraphics[width=1\linewidth]{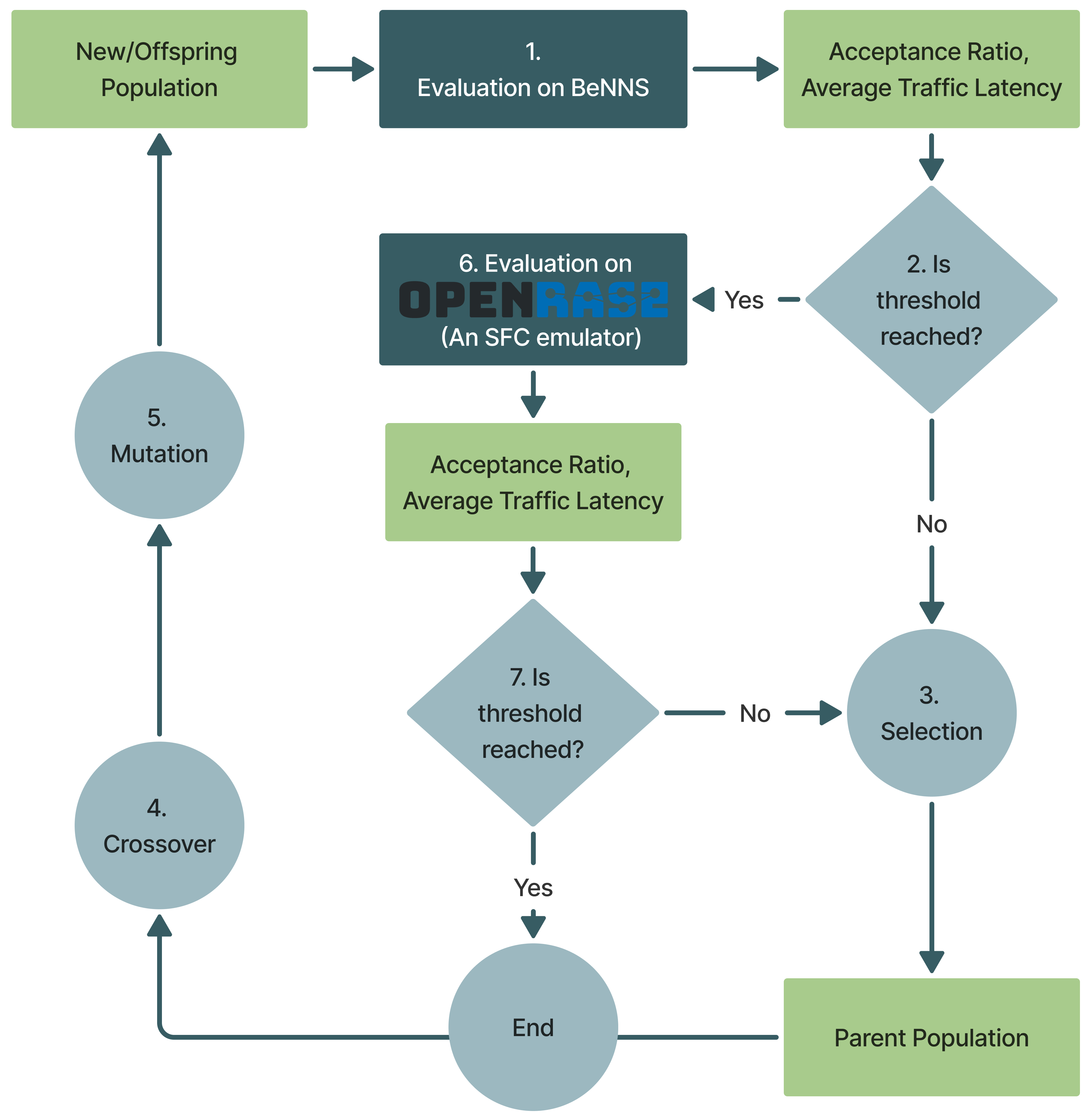}
    \caption{The hybrid online-offline evolution approach}
    \label{fig:hybrid}
\end{figure}

In Step 1 in Fig.~\ref{fig:hybrid}, we start with approximating the fitness of an initial \textit{population} of \textit{individuals}--each representing a different solution to the OSE problem--offline using BENNS. In Step 2, we check if the acceptance ratio and the average traffic latency are within the threshold. The threshold is defined in terms of a minimum acceptance ratio and a maximum average traffic latency based on the business needs of the cloud service provider and the Service Level Agreement with the clients.
If not, we move to Step 3, where we select the parents using a selection strategy such as NSGA II, tournament selection, roulette wheel selection, etc. 
In Step 4, we perform crossover between individuals in the parent population.
In Step 5, we mutate the individuals in the parent population. 
After this step, we end up with an offspring population in addition to the parent population. We then evaluate the offspring on BENNS and iterate over Steps 1-5 until the threshold is met in Step 2.
If at least one individual meets the threshold in Step 2, we move to Step 6 and evaluate that individual online on OpenRASE.
If the individual meets the threshold in Step 7 after the online evaluation, an optimal individual has been found, and evolution stops. If not, which can happen if there is a discrepancy between fitness approximated by BENNS and measured on OpenRASE, we update the approximated fitness of the individual with the fitness measured on OpenRASE and proceed to Step 3, i.e. parent selection. In the optimal case, the average of the traffic latency approximated by BENNS is comparable to the average of the measured traffic latency, meaning the online evolution becomes a verification step. 

Unlike other approaches, in the hybrid approach, the fitness of an optimal solution is both approximated offline and measured online. Online evaluation is not only for validation, but also for course correction if there is going to be a discrepancy between the approximation and the measurement. The approach may go back and forth between offline and online evaluation, ensuring the evolution gravitates toward the optimal region in the search space. 

\section{BENNS Model}
\label{sec:benns}
\begin{figure*}
    \Description{The architecture of BENNS}
    \centering
    \includegraphics[width=1\linewidth]{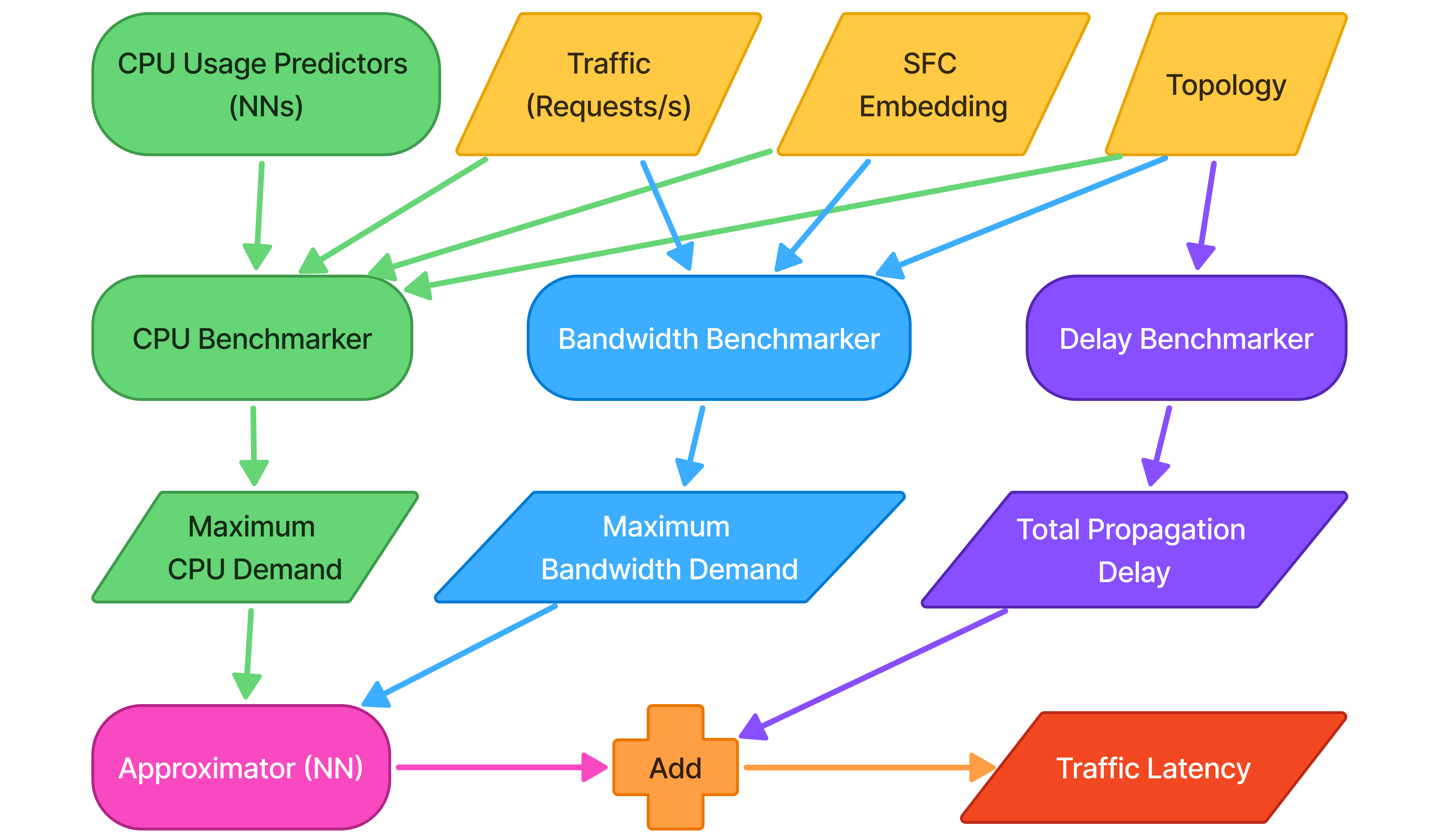}
    \caption{A diagram showing how BENNS approximates traffic latency}
    \label{fig:benns}
\end{figure*}

To enable reasonable accuracy and efficient offline evaluation in our approach, we developed a surrogate model called BENNS to approximate the fitness of GA-generated solutions to the OSE problem, as shown in Fig.~\ref{fig:benns}. The objective of the surrogate is not to predict the fitness accurately but to approximate it \emph{faster}, ensuring unfit individuals are not evaluated online, and fit individuals are not rejected, saving time. Consequently, we wanted to develop a simple and lightweight approximation model to enable faster approximation of fitness.

\subsection{Solution Overview}
The surrogate model is based on benchmarking models and a pre-trained NN, which we call an \textit{approximator}, that approximates the traffic latency. Given the complexity of SFC embeddings and network topologies, we first benchmark the SFC embedding against the network topology, encoding them to numerical values, and use these numerical values as inputs to the approximator. Based on assumptions taken by existing works in the literature~\cite{Liu2021a, Gamal2019a, Fu2020, Tahmasebi2018, Liu2024a, Khoshkholghi2020}, we assumed that the traffic latency of an SFC is a function of the maximum CPU and memory demanded of the hosts that host the VNFs of an SFC, and the maximum bandwidth demanded of the physical links connecting the VNFs of the SFC. The maximum bandwidth demand is calculated considering the number of HTTP requests an SFC receives. To calculate the maximum CPU and memory demand, we use CPU and memory usage predictors for each VNF, which are NNs trained on CPU and memory usage data of each VNF generated by profiling it (Section~\ref{predict_profiling}). This factors in the processing heterogeneity of VNFs. We performed an empirical analysis (Section~\ref{toy_exp}) of our assumptions on resource demands and found that only the maximum CPU demand and bandwidth demand have an impact on the traffic latency (Section~\ref{empirical_analysis_results}). 

We then generated data to train the approximator by running 11 experiments across different topologies and SFC embeddings using a dynamic traffic pattern (Section~\ref{apprx_dg}). We computed the maximum CPU demand and bandwidth demand, and the total propagation delay of every SFC for a given number of requests, and recorded them against the traffic latency measured on OpenRASE. This data captured the contention and multiplexing effects of VNFs on shared hosts and physical links, and their impact on traffic latency. We used this data to train the approximator. To approximate the traffic latency of an SFC for a given number of requests, we compute the maximum CPU and bandwidth demand and input it to the approximator and have it approximate the traffic latency. We compute the total propagation delay of SFCs and add it to the approximated traffic latency for reasons discussed in Section~\ref{approx}. We average the approximated traffic latency for all SFCs across the entire duration of traffic to produce the average traffic latency of an SFC embedding. 

\subsection{Benchmarking SFCs}\label{benchmark}
Approximating the fitness of an SFC embedding for a given topology and traffic flow is challenging because it is a complex graph consisting of several dimensions. At the same time, the SFC embedding and its network topology have to be encoded into numerical values so that they can be input to the approximator, which is an NN. Thus, benchmarking SFCs in their network topology serves two purposes:
\begin{enumerate*}
    \item \textbf{dimensionality reduction}, by reducing a high-dimensional SFC embedding and its network topology to two dimensions, and
    \item \textbf{numerical encoding}, by converting them to numerical values that can be input to the approximator, which is an NN.
\end{enumerate*} 
This approach allows BENNS to be effective across different SFC embeddings, traffic patterns and network topologies (Section~\ref{results_hybrid}). In other words, BENNS is SFC embedding, network topology and traffic agnostic. 

We first assumed that the performance of an SFC embedding in its network topology is determined by the 
\begin{enumerate*}
    \item maximum CPU demanded of hosts
    \item maximum memory demanded of hosts
    \item maximum bandwidth demanded of physical links.
\end{enumerate*}. These assumptions are based on the assumptions used for numerical computation of SFC performance in the literature~\cite{Liu2021a, Gamal2019a, Fu2020, Tahmasebi2018, Liu2024a, Khoshkholghi2020}. We developed a mathematical model (Section~\ref{math_model}), and CPU and memory usage predictors to compute the maximum CPU, memory and bandwidth demands of SFCs. We validated these assumptions empirically (Section~\ref{toy_exp}) by studying the impact of each resource demand on traffic latency, unlike existing works in the literature that\emph{ did not validate their assumptions.} 

\subsubsection{Mathematical Modelling}\label{math_model}
\begin{table}
    \caption{Variables used in mathematical modelling.}    
    \begin{tabular}{p{0.15\linewidth} p{0.75\linewidth}}
        \hline
         \thead{\textbf{Variable}}& \thead{\textbf{Description}}\\ \hline
         $c_h$& no. of CPUs available in host $h$\\
         $m_h$& amount of memory available in host $h$\\
         $b_l$& bandwidth of physical link $l$\\
 $H(s)$&set of hosts that host at least one VNF of SFC $s$\\
 $S(l)$&set of SFCs that use physical link $l$\\
 $S(h)$&set of SFCs that use host $h$\\
 $Y(s,l)$&set of virtual links of SFC $s$ that use physical link $l$\\
 $V(s,h)$&set of VNFs of SFC $s$ hosted in host $h$\\
 $L(s)$&set of physical links used by SFC $s$\\
 $j_{ys}$&level of virtual link $y$ in its SFC $s$\\
 $k_{vs}$&level of VNF $v$ in its SFC $s$\\
          $r_{st}$& no. of requests being processed by SFC $s$ at time slot $t$\\
 $d_l$&propagation delay of physical link $l$\\
\hline
 
    \end{tabular}
    \label{tab:var}
\end{table}
As mentioned above, we assume that the maximum CPU and memory demanded of hosts, as well as the bandwidth demanded of the physical links, have an impact on the performance of SFCs based on existing works~\cite{Liu2021a, Gamal2019a, Fu2020, Tahmasebi2018, Liu2024a, Khoshkholghi2020}. This is also based on the intuition that hosts with the maximum CPU and memory demands, and physical links with the maximum bandwidth demand, may become performance bottlenecks. Also, the propagation delay of each physical link, which depends on the physical nature of the link, impacts the traffic latency. CPU and memory demands are functions of the number of requests processed by VNFs hosted in a host (Section~\ref{predict_profiling}), and bandwidth demand is a function of the number of requests travelling via a physical link. A physical link is a physical connection among hosts and switches in the topology. In contrast, a virtual link is a link between successive VNFs in an SFC. A virtual link is embedded over multiple physical links. The duration of an experiment is broken into time slots, and the demands are calculated for each time slot. 

The CPU and memory demand of a host is the sum of the CPU and memory demands of the VNFs deployed in it, divided by the total CPU and memory available in the host. The CPU and memory demands of a VNF depend on the number of requests being processed by the VNF. The number of requests processed by a VNF depends on the number of requests received by its SFC and its level within the SFC. An SFC may end up with multiple levels if one of its VNFs is a traffic splitter, such as a load balancer. In this study, we only consider traffic splitters that split traffic into two branches. An SFC with one traffic splitter has two levels, with virtual links and VNFs in the second level receiving half of the requests to the SFC. To compute the CPU and memory usages of a VNF for a given number of requests, the $CpuPred$ and $MemPred$ functions are used. Section~\ref{cpu_mem_pred} discusses how these functions work.

All terms and symbols used in the following are described in Table~\ref{tab:var}. The CPU demand ($f_{cpu}$) of a host $h$ at a time slot $t$ is given by the sum of the CPUs demanded by every VNF deployed in the host, divided by the number of CPUs available in that host. CPUs demanded by VNFs are a function of the number of requests received by its SFC and its level within the SFC. 
 \begin{align*}
f_{cpu}(h,t) = \frac{\sum_{s \in S(h)}{\sum_{v \in V(s,h)} {CpuPred(v, \frac{r_{st}}{2^{k_{vs}-1}}}})}{c_h}
 \end{align*}

Similarly, the memory demand ($f_{memory}$) of a host $h$ at a time slot $t$ is given by:
 \begin{align*}
f_{memory}(h,t) = \frac{\sum_{s \in S(h)}{\sum_{v \in V(s,h)}  {MemPred(v, \frac{r_{st}}{2^{k_{vs}-1}}}})}{m_h}
 \end{align*}
The bandwidth demand ($f_{bw}$) of a physical link $l$ at a time slot $t$ is the sum of the requests processed by all virtual links embedded upon that physical link, divided by the bandwidth of the physical link. The number of requests processed by a virtual link is a function of the number of requests received by its SFC and its level within the SFC.
\begin{align*}
f_{bw}(l, t) = \frac{\sum_{s \in S(l)}\sum_{y \in Y(s,l)}{\frac{r_{st}}{2^{j_{ys}-1}}}}{b_l}
\end{align*}

The \textit{maximum CPU demand} of an SFC $s$ at a time slot $t$ is the maximum CPU demanded of all hosts hosting at least one VNF in SFC $s$.
\begin{equation}\label{eq:cpu_score}
\alpha(s,t) = max(\set{p \mid p = f_{cpu}(h, t)\text{ and }h \in H(s)})
\end{equation}
Similarly, the \textit{maximum memory demand} of an SFC $s$ at a time slot $t$ is the maximum memory demanded of all hosts hosting at least one VNF in SFC $s$.
\begin{equation}\label{eq:mem_score}
\gamma(s,t) = max(\set{e \mid e = f_{memory}(h, t)\text{ and }h \in H(s)})
\end{equation}
The \textit{maximum bandwidth demand} of an SFC $s$ at a time slot $t$ is the maximum bandwidth demanded of all the physical links used by SFC $s$.
\begin{equation}\label{eq:link_score}
    \beta(s,t) = max(\set{z \mid z = f_{bw}(l,t) \text{ and } l \in L(s)})
\end{equation}
The \textit{total propagation delay} of an SFC $s$ is the sum of the propagation delay of every virtual link in an SFC $s$. We multiply this by $2$ to factor in the propagation delay during response. Summing the propagation delays of virtual links also factors in the delay introduced by the hop count. 
\begin{equation}\label{eq:delay}
    \delta(s) = 2 \times (\sum_{l \in L(s)}{\sum_{y \in Y(s,l)}{\frac{d_l}{2^{j_{ys}-1}}}})
\end{equation}
\subsubsection{CPU and Memory Usage Prediction}\label{cpu_mem_pred}
The CPU and memory usage of VNFs is predicted based on the number of requests the VNF has to process at that given time. OpenRASE provides 7 VNF implementations (Table~\ref{tab:vnf_nn}) that are deployed in Docker containers~\cite{OpenRASE}. We developed CPU and memory usage predictors per VNF, which are NN models, to predict the CPU and memory usage of each VNF, unlike other offline approaches that require these to be user-specified.

The CPU and memory usage predictors are fully-connected, feed-forward NNs trained on profiling data for each VNF and resource type to predict the CPU and memory usage of each VNF for a given number of requests.  Section~\ref{predict_profiling} describes how VNFs were profiled to generate data to train the usage predictors. The architecture of the usage predictors for each VNF and resource type is given in Table~\ref{tab:vnf_nn}. Each predictor uses the ReLU activation function and He Normal initialisation. 
\begin{table}
    \caption{The architecture of CPU and memory usage predictors is given by the number of neurons in each layer. E.g., 1, 32, 64, 1 means that 1, 32, 64 and 1 neurons are in the input, first-hidden, second-hidden, and output layers, respectively. }
    \begin{tabular}{p{0.4\linewidth}p{0.2\linewidth}p{0.3\linewidth}} 
         \hline
         \thead{\textbf{VNF}}&  \thead{\textbf{CPU Demand}}&  \thead{\textbf{Memory Demand}}\\
         \hline
         Web Application Firewall&  1, 32, 32, 1&  1, 32, 32, 64, 1\\
         Load Balancer&  1, 4, 24, 1&  1, 32, 64, 64, 128, 1\\
         Intrusion Detection System&  1, 8, 32, 32, 1&  1, 64, 64, 256, 1\\
         Intrusion Prevention System&  1, 8, 16, 16, 1&  1, 64, 64, 128, 1\\
         Traffic Monitor&  1, 16, 16, 32, 1&  1, 64, 64, 64, 1\\
         Deep Packet Inspection&  1, 4, 4, 1&  1, 64, 128, 128, 1\\
 Http Accelerator& 1, 64, 64, 128, 1& 1, 150, 150, 256, 1\\
 \hline
    \end{tabular}
    
    \label{tab:vnf_nn}
\end{table}

The $CpuPred$ and $MemPred$ functions take the VNF type and the number of requests as inputs, call the relevant usage predictor and output the predicted CPU and memory demands, respectively. 

\subsection{Average Traffic Latency Approximation}\label{approx}
To approximate the average traffic latency, we developed an approximator, which is a fully-connected, feed-forward NN with 2 hidden layers with 16 neurons each, using the Sigmoid activation function, as it performed the best in comparison to Support Vector Regression (SVR), Gaussian Process Regression (GPR), Linear Regression (LR) and other NN architectures (Section~\ref{result_eval_approx}). Glorot Normal was used for parameter initialisation, and Adamax was used for parameter optimisation with a learning rate of 0.05. The empirical analysis of the assumptions made (Section~\ref{empirical_analysis_results}) showed that only the maximum CPU and bandwidth demands impacted the traffic latency. Therefore, the approximator was trained to approximate the traffic latency for a given maximum CPU and bandwidth demands. Section~\ref{apprx_dg} describes how the data was generated to train the approximator. The simple, lightweight architecture of the NN enables faster approximation (Section~\ref{results_hybrid}) even on CPUs, obviating the need for GPUs (Section~\ref{exp_setup}). 

Before training, outliers were removed from the dataset using the Inter-Quartile Range method. We subtracted the total propagation delay from the traffic latencies because the total propagation delay for an SFC remains constant irrespective of the traffic flow, and the traffic latency is directly proportional to the total propagation delay. Subtracting the total propagation delay from the traffic latency allows the approximator to learn the proportional influence of the maximum CPU and bandwidth demands on the traffic latency. To reduce noise, 20 successive time slots were binned, and the median maximum CPU and bandwidth demands, and traffic latencies were computed. 70\% of the data was used for training, with 10\% used for validation and 20\% for testing. The data was first normalised and then trained over 200 epochs. Evaluation results of the approximator are given in Section~\ref{result_eval_approx}.

To use the approximator, for a given SFC embedding and network topology, we first compute the maximum CPU (Equation~\ref{eq:cpu_score}) and bandwidth demands (Equation~\ref{eq:link_score}) and the total propagation delay (Equation~\ref{eq:delay}) for each SFC and for each time slot during the experiment. We then input the maximum CPU and memory demands to the approximator to approximate the traffic latency. We then add the total propagation delay to the traffic latencies and average them across all SFCs and time slots to compute the average traffic latency of the given SFC embedding. We add the total propagation delay to the traffic latency approximated by the approximator because we subtract the total propagation delay from the traffic latency when training the approximator.

\section{Evaluation}
We describe four sets of experiments used to generate data to train the CPU and memory usage predictors, empirically analyse the assumptions we made in Section~\ref{benchmark}, generate data to train the approximator, and validate the hybrid online-offline approach to optimising the OSE problem. 

\subsection{Experimental Setup}\label{exp_setup}
To carry out the experiments, we used a test machine, which is a virtual machine running Ubuntu 20.04.6 on a QEMU hypervisor with 64 cores of Intel Xeon Gold 6240R CPUs having a clock speed of 2.4 GHz, and 64GB of RAM. 

We chose a data centre environment as the use case, as it is a common use case for OSE problems~\cite{Xie2021, Jang2017, Xu2021, Wang2020a}. A 4-ary fat-tree topology was chosen as the substrate network since fat-tree topologies are a common network topology used in data centres~\cite{8647774}.

Four SFCRs based on VNFs used in data centres~\cite{Yang2016Energy-awareCenters, Herker2015Data-centerRequirements, Zu2021FairCenter} were used in the experiments, and these SFCRs are as follows:
\begin{enumerate*}
    \item Load Balancer $\rightarrow $ Web Application Firewall
    \item HTTP Accelerator $\rightarrow$ Load Balancer $\rightarrow$ Web Application Firewall
    \item HTTP Accelerator $\rightarrow$ Traffic Monitor $\rightarrow$ Load Balancer $\rightarrow$ Web Application Firewall
    \item Load Balancer $\rightarrow$ Traffic Monitor $\rightarrow$ Web Application Firewall 
\end{enumerate*}.

We set the time slot to 1s (Section~\ref{math_model}) for all experiments carried out in this study. The CPUs and memory available to hosts and the bandwidth of the physical links were set to smaller values (Table~\ref{tab:experiments} and Table~\ref{tab:dg_exp}) to emulate a miniaturised data centre environment, as emulating the real resource availability of a data centre was impractical in the test machine. 

\subsection{Profiling VNFs}\label{predict_profiling}
To generate data to train the CPU and memory usage predictors per VNF, each VNF was deployed in a host with access to all the CPUs and memory available in the test machine running OpenRASE, and traffic was generated in the form of HTTP requests. The profiling lasted for 5 minutes, and the number of requests was increased from 1 request per second to 500 requests per second linearly across the 5 minutes. Every 2s, the number of CPUs and the amount of memory used in MB were recorded along with the number of requests being processed. The 2s window was chosen because that is the shortest time window OpenRASE provides to measure resource usage. We then used this data to train the CPU and memory usage predictors for each VNF as described in Section~\ref{cpu_mem_pred}. Profiling each VNF individually enables BENNS to capture the heterogeneous processing and memory requirements of VNFs. 

\subsection{Empirical Analysis of Resource-Demand Assumptions}\label{toy_exp}
As described in Section~\ref{benchmark}, we assumed that the traffic latency of an SFC was affected by the maximum CPU, memory and bandwidth demands. We carried out an empirical analysis to validate these assumptions. We conducted three experiments to test the assumptions on the impact of the maximum CPU, memory and bandwidth demands. Each experiment evaluated the impact of a resource demand, and only the relevant resource was constrained. The baseline network resource configuration was set to 64 CPUs and 64GB of memory per host, while the bandwidth was left unrestricted. The propagation delay of all physical links was set to 1ms. To test the maximum CPU demand, the CPUs available to each host were restricted to 0.5. To test the maximum memory demand, the memory was restricted to 1GB, while the link bandwidth was restricted to 4MB to test the bandwidth demand. Since the CPU, memory and bandwidth demands are a function of the resource requested divided by the resource available (Section~\ref{math_model}), by constraining the resource available, we are increasing the demand. 

We used the first two SFCRs listed in Section~\ref{exp_setup} for the empirical analysis. To study the impact of maximum CPU and memory demands on traffic latency, all SFCs were deployed in one host to further increase the maximum CPU and memory demands. The SFCRs were deployed in random hosts to study the impact of the maximum bandwidth demand. Each experiment consisted of 5 rounds, with the traffic being linearly ramped up to 250 requests per second from 0 across 2 minutes. Linearly ramping up the traffic gradually increases the resource demand, allowing us to analyse how the traffic latency changes as the demand increases.  

The maximum CPU, memory and bandwidth demands and the total propagation delay for each SFC at each 1s timeslot were calculated using Equations~\ref{eq:cpu_score},~\ref{eq:mem_score}, and~\ref{eq:link_score}. The traffic latencies during those timeslots were measured and recorded during the experiment, and the total propagation delay was subtracted from the traffic latency for reasons discussed in Section~\ref{approx}. The results of these experiments are provided in Section~\ref{empirical_analysis_results}.

\subsection{Generating Data for Approximator}\label{apprx_dg}
The data for pre-training the approximator was generated by performing 11 experiments on OpenRASE, each with a different topology or traffic pattern, as shown in Table~\ref{tab:experiments}. The memory available to a host and the propagation delay of links were kept at 5GB and 1ms for all experiments, while the CPUs available in a host, link bandwidth and traffic patterns were altered. 

8 copies of each SFCR listed in Section~\ref{exp_setup} ($4\times 8=32$) were used to generate 20 random SFC embeddings for each experiment. Thus, altogether, $11 \times 20 = 220$ evaluations were carried out to generate data.
\begin{table}
    \caption{The configurations of experiments for generating data to train the approximator.}
    \begin{tabular}{p{0.19\linewidth}p{0.11\linewidth}p{0.15\linewidth}p{0.37\linewidth}}
            \hline
           \thead{\textbf{Experiment}}&  \thead{\textbf{CPUs}\\ \textbf{per host}}& \thead{\textbf{Link}\\ \textbf{Bandwidth}}&\thead{\textbf{Traffic Pattern}} \\
            \hline
           1&  1& 5Mbps&1req/s$\rightarrow$50req/s in 300s\\
           2&  0.5& 20Mbps&1req/s$\rightarrow$50req/s in 300s\\
           3&  0.5& 5Mbps&1req/s$\rightarrow$50req/s in 300s\\
           4&  1& 20Mbps&1req/s$\rightarrow$50req/s in 300s\\
           5& 4& 100Mbps&1req/s$\rightarrow$25req/s in 60s\\
           6& 0.2& 50Mbps&20req/s$\rightarrow$25req/s in 30s\\
           7& 0.2& 5Mbps&20req/s$\rightarrow$25req/s in 30s\\
           8& 1& 50Mbps&20req/s$\rightarrow$25req/s in 30s\\
           9& 0.5& 5Mbps&20req/s$\rightarrow$25req/s in 30s\\
           10& 0.2& 20Mbps&20req/s$\rightarrow$25req/s in 30s\\
           11& 1& 20Mbps&20req/s$\rightarrow$25req/s in 30s\\
 \hline
    \end{tabular}
    \label{tab:dg_exp}
\end{table}

During the experiments, the maximum CPU (Equation~\ref{eq:cpu_score}) and bandwidth (Equation~\ref{eq:link_score}) demands of SFCs across 1s timeslots, and the total propagation delay (Equation~\ref{eq:delay}) of SFCs were recorded along with the measured traffic latency.  

\subsection{Evaluation of Hybrid Online-Offline Evolution}\label{eval_hybrid}
\begin{table}
    \caption{The configurations of experiments for evaluating hybrid evolution.}
    \begin{tabular}{p{0.18\linewidth}p{0.11\linewidth}p{0.15\linewidth}p{0.10\linewidth}p{0.09\linewidth}p{0.10\linewidth}}
        \hline
         \thead{\textbf{Experiment}}&\thead{\textbf{CPUs}\\ \textbf{per host}}&\thead{\textbf{Link}\\ \textbf{Bandwidth}}&\thead{\textbf{No. of}\\ \textbf{SFCRs}}&\thead{\textbf{Traffic}\\ \textbf{Scale}}&\thead{\textbf{Traffic}\\ \textbf{Pattern}}\\
         \hline
         Basic&  2&  10Mbps&  32&  1& A\\
         CPU&  1&  10Mbps&  32&  1& A\\ 
         Bandwidth&  2&  5Mbps&  32&  1& A\\
         Traffic Scale&  2&  10Mbps&  32&  2& A\\
         Traffic Pattern&  2&  10Mbps&  32&  1& B\\
        \hline
    \end{tabular}
    
    \label{tab:experiments}
\end{table}
To demonstrate the effectiveness and feasibility of the hybrid online-offline approach using BENNS to optimise the OSE problem across diverse network scenarios, we carried out experiments under different traffic flows and topological configurations. 

\paragraph{Baselines}
We compare our hybrid approach to the online Binary Encoded Genetic Algorithm (BEGA Online)~\cite{OpenRASE}, and offline Genetic Algorithm-based Heuristic Approach (GAHA Offline)~\cite{Khoshkholghi2020} , and a non-GA Greedy Dijkstra Algorithm (GDA)\footnote{\url{https://sourceforge.net/p/alevin/svn/HEAD/tree/trunk/src/vnreal/algorithms/samples/SimpleDijkstraAlgorithm.java}} from the literature. BEGA Online was shown to be of higher accuracy than BEGA Offline, but at the cost of increased time. GAHA mathematically computes the traffic latency. 

\paragraph{Optimisation Problem}
We adapt BEGA Online to evaluate our hybrid approach, and we refer to it as BEGA Hybrid henceforth. BEGA Online and GAHA Offline optimised only the VNF embedding and link embedding sub-problems of the OSE problem and considered only a static network scenario, where the number of SFCRs to embed, network topology and traffic remained constant. To demonstrate our hybrid approach's effectiveness in optimising all three OSE sub-problems, we evolve three neural networks using GA to optimise all three OSE sub-problems using the hybrid approach, named GENESIS Hybrid. We then evaluate the hybrid approach's effectiveness, using GENESIS Hybrid, in a dynamic network environment where every 5 minutes, a set of 4 SFCRs arrives, and the traffic pattern changes. During the second half of the experiment, every 5 minutes, a host crashes, increasing the complexity of the problem.

\paragraph{GA Configuration}
For GAs, the threshold was set to a \textit{minimum acceptance ratio of 1} and a \textit{maximum average traffic latency of 150ms}. In addition, BEGA Online was stopped after 10 generations and hybrid evolution, GAHA Offline and GENESIS Hybrid were stopped after 500 generations if they failed to produce an individual that met the threshold. The cap on the number of generations serves as a time limit. The limit was set to 10 for BEGA Online and 500 for others, as online evolution takes longer. For all evolutions, the initial population was randomly generated from a uniform distribution. For individuals with an acceptance ratio of 0, the average traffic latency was set to a very high value of 50000 to eliminate them from the population.

\paragraph{Experiment Traffic}
The traffic pattern used for the experiments was adapted from a university campus data centre~\cite{Benson2010NetworkWild}. The traffic pattern contains the number of requests per hour over a week. To vary the traffic, we increased the amount of traffic by multiplying the number of requests by two (Traffic Scale experiment), and produced a different traffic pattern by phase shifting the traffic by 50\% (Traffic Pattern experiment). Different topological configurations were created by altering the resources available in a topology, such as the number of CPUs available in a host and the bandwidth of links. All told, 6 different experiments were designed as shown in Table~\ref{tab:experiments}. The Basic experiments were the least demanding of all experiments, as neither the resources were constrained nor the demand was high. All other experiments increased the complexity by either constraining resources or increasing the demand.

\section{Results}\label{results}
This section discusses the results of the empirical analysis (Section~\ref{toy_exp}), which informed the design of the BENNS model, the evaluation of the approximator (Section~\ref{approx}), and the evaluation of the hybrid evolution (Section~\ref{eval_hybrid}). 
\subsection{Results of Empirical Analysis}\label{empirical_analysis_results}
\begin{figure}
\Description{Scatter plots showing the impact of maximum CPU, memory and bandwidth demands on traffic latency}
    \centering
    \includegraphics[width=1\linewidth]{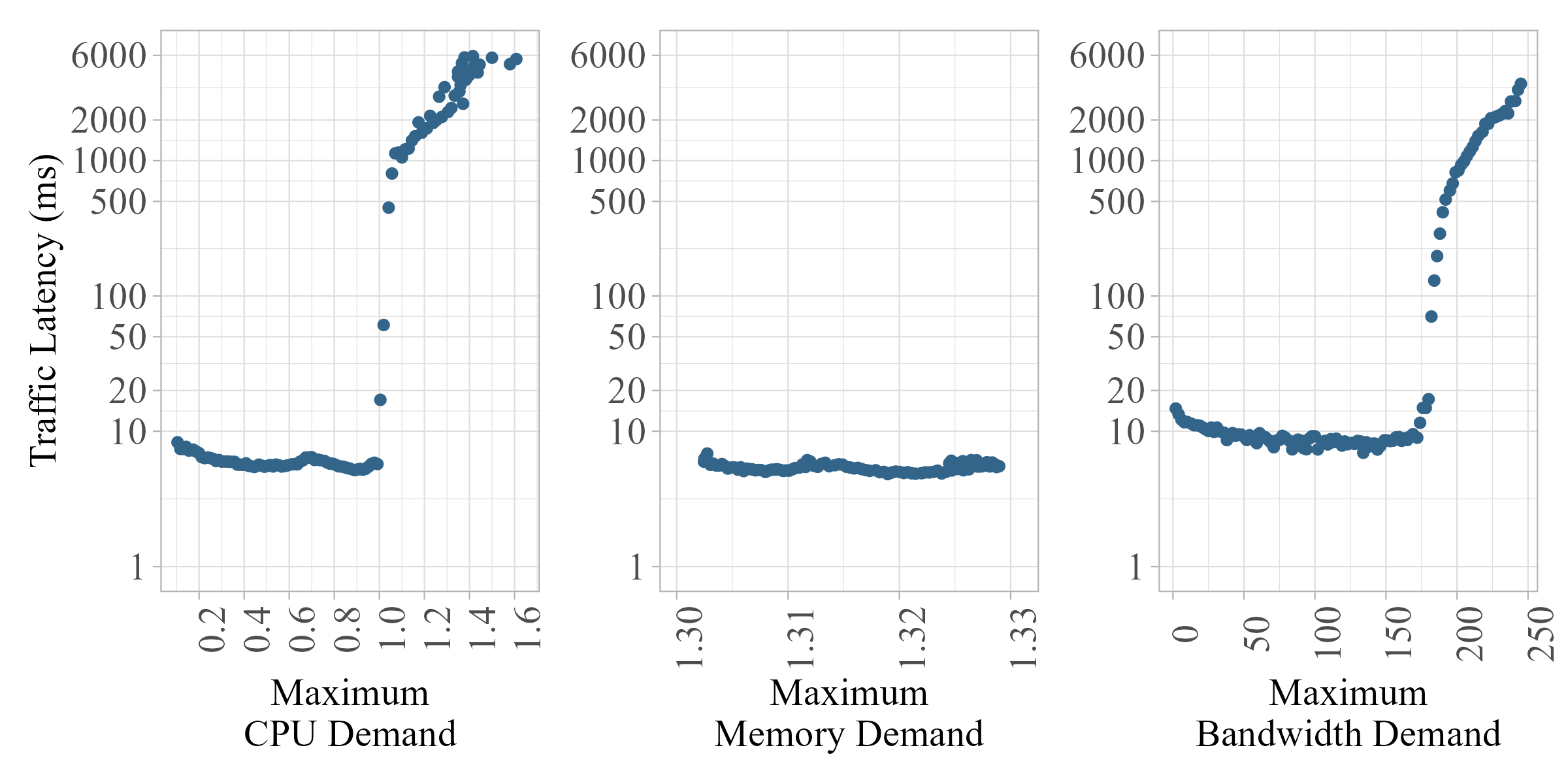}
    \caption{Scatter plots showing the relationship between traffic latency and CPU, memory and bandwidth demand, respectively}
    \label{fig:toy_exp}
\end{figure}

Fig.~\ref{fig:toy_exp} shows three scatter plots drawn for each of the three experiments carried out to analyse the impact of the maximum CPU, memory and bandwidth demands on the traffic latency. It can be seen that the maximum CPU demand and the maximum bandwidth demand have an impact on the measured traffic latency, whereas the maximum memory demand does not have an impact. As the maximum CPU demand got close to 1, implying CPUs requested exceeded CPUs available (Section~\ref{math_model}), the traffic latency increased sharply. Similarly, as the maximum bandwidth demand approached 175, the traffic latency rose sharply. However, traffic latency remained constant even though the maximum memory demand exceeded 1.

\subsection{Results of Evaluation of Approximator Models}\label{result_eval_approx}
The approximator models were evaluated on the test dataset to find their approximation accuracy. The models included SVR, LR, GPR, and NNs with different architectures, as shown in Table~\ref{tab:model_perf}. The NN consisting of two hidden layers with 16 neurons each using the Sigmoid activation function produced the best Mean Absolute Percentage Error (MAPE) of 0.86 and was chosen as the model of the approximator. 

\begin{table}
    \centering
        \caption{The MAPE of the models used to approximate the average traffic latency. The number of hidden layers, neurons and the activation function of the NNs are described similarly to Table~\ref{tab:vnf_nn}.  }\begin{tabular}{p{0.1\linewidth}p{0.7\linewidth}p{0.1\linewidth}}\toprule
         \thead{\textbf{Model}}&  \thead{\textbf{Parametres}}&\thead{\textbf{MAPE}}\\\midrule
         SVR&  RBF kernel&5.17\\
         LR&  &21.47\\
         GPR&  Fixed RBF kernel&1.08\\
         NN&  16,16 (Sigmoid)&0.86\\
         NN&  16 (Sigmoid)&1.06\\
         NN&  16,16,16 (Sigmoid)&1.08\\
         NN&  32 (Sigmoid)&1.18\\
         NN&  16, 16 (ReLU)&2.73\\
 NN& 16,16 (linear function)&21.27\\ \bottomrule
    \end{tabular}

    \label{tab:model_perf}
\end{table}

\begin{figure}
    \Description{The fitness landscape of the OSE problem}
    \centering
    \includegraphics[width=1\linewidth]{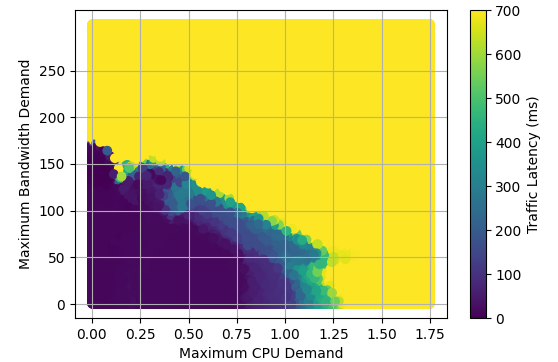}
    \caption{The fitness landscape of the OSE problem as approximated by BENNS}
    \label{fig:fitness_landscape}
\end{figure}

To visualise the fitness landscape as approximated by BENNS, we generated 1 million random maximum CPU and bandwidth demands and had BENNS approximate their traffic latencies. We plotted a scatter plot with the colour showing the traffic latency as shown in Fig.~\ref{fig:fitness_landscape}.

\subsection{Results of Hybrid Online-Offline Evolution}\label{results_hybrid}
\begin{figure*}
\Description{A column chart showing the time taken by the different approaches}
    \centering
    \includegraphics[width=1\linewidth]{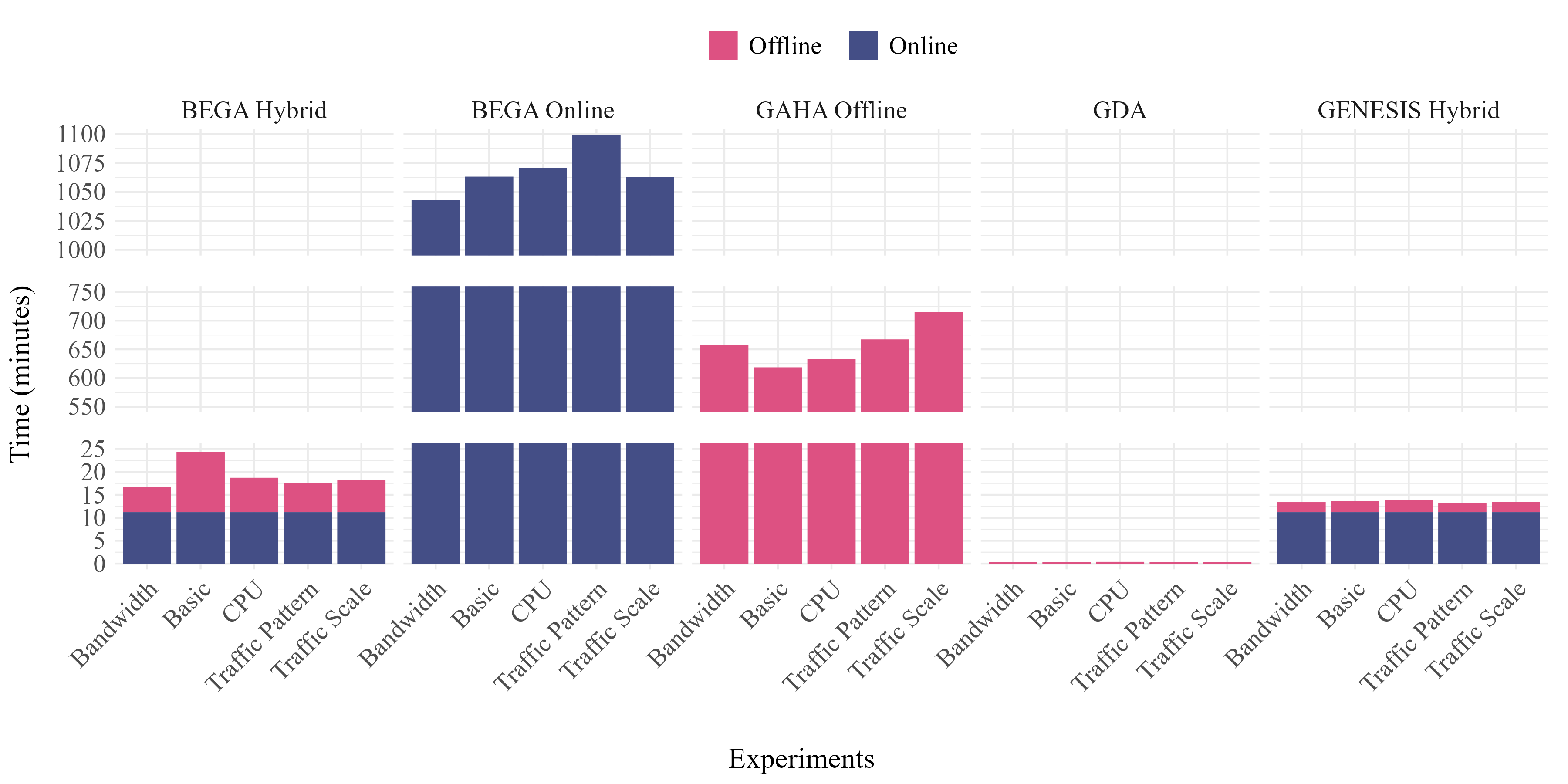}
    \caption{The time taken by online and hybrid experiments}
    \label{fig:time_taken}
\end{figure*}

\begin{figure}
    \centering
    \Description{Line plots showing how the acceptance ratio changed over generations}
    \includegraphics[width=1\linewidth]{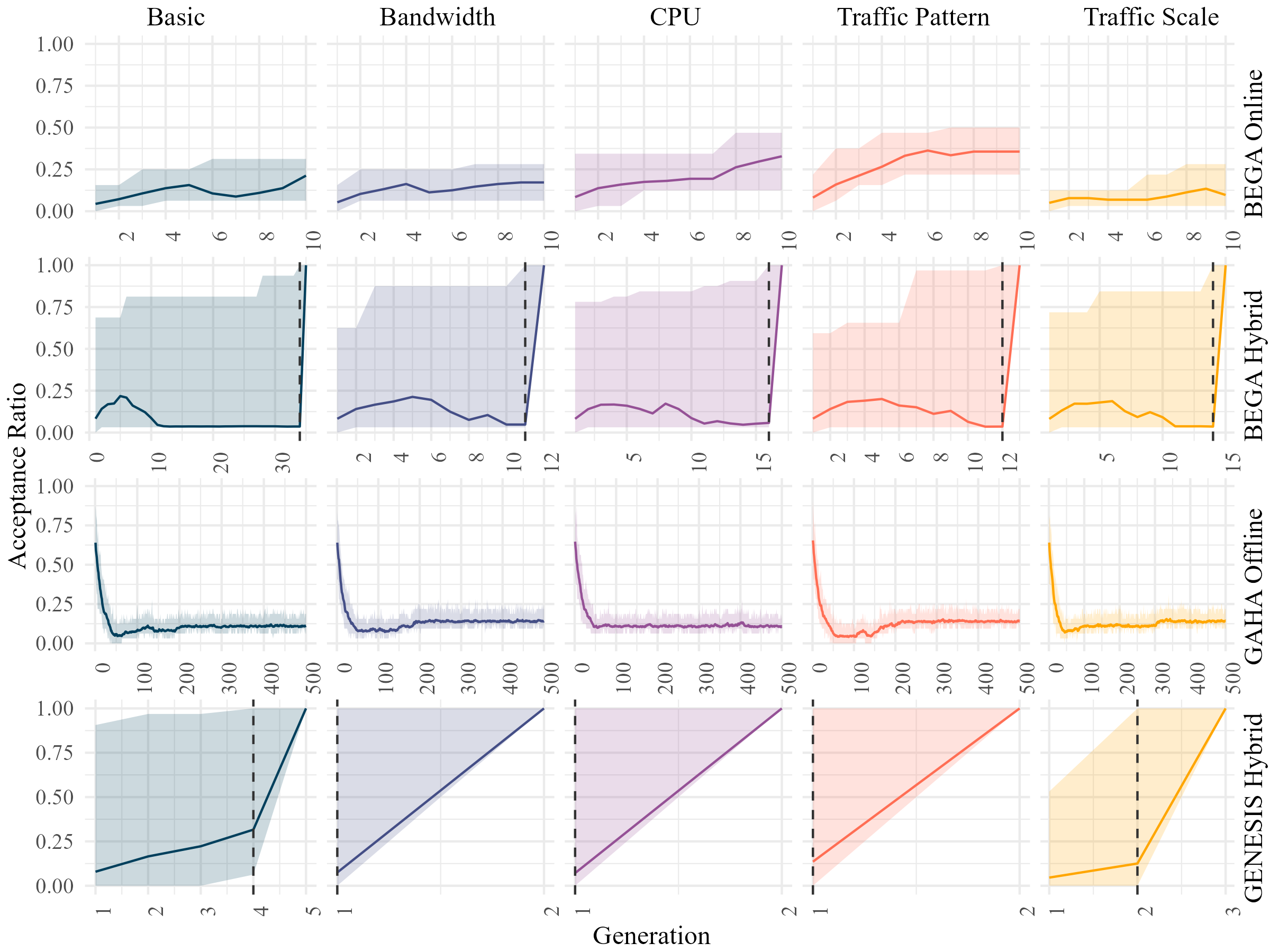}
    \caption{The acceptance ratios by generation of experiments. The curves show the mean acceptance ratio of individuals in a generation, while the shaded area covers the maximum and minimum acceptance ratios. The dashed line shows the point at which the transition from offline to online evolution took place.}
    \label{fig:ar_plot}
\end{figure}

\begin{figure}
    \centering
    \Description{Line plots showing how the average tarffic latency changed over generations}
    \includegraphics[width=1\linewidth]{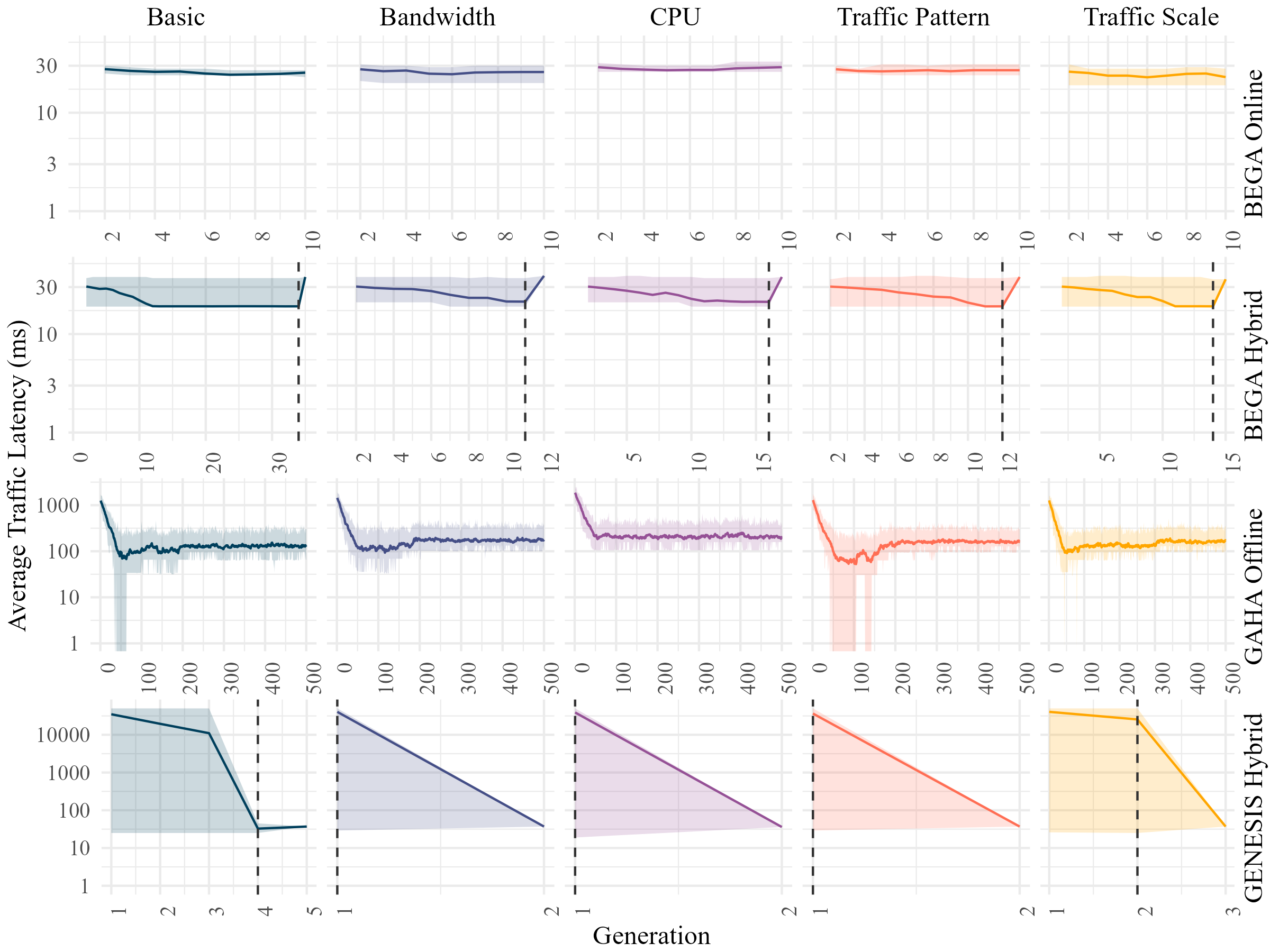}
    \caption{The average traffic latencies by generation of experiments. The curves show the mean average traffic latency of individuals in a generation, while the shaded area covers the maximum and minimum traffic latencies. The dashed line shows the point at which the transition from offline to online evolution took place.}
    \label{fig:traffic_latency}
\end{figure}

GDA was the fastest algorithm, but it failed to converge in any experiment. GAHA Offline, which could not converge within 500 generations in any experiment, was slower than BEGA Hybrid and GENESIS Hybrid, but was faster than BEGA Online. The hybrid evolutions converged faster than online evolutions, which failed to converge, as shown in Fig.~\ref{fig:time_taken}. The change in acceptance ratios by generation for all experiments is given in Fig.~\ref{fig:ar_plot}, while Fig.~\ref{fig:traffic_latency} shows the change in average traffic latencies across generations.

For all 6 experiments, the BEGA Online and GAHA Offline failed to meet the threshold within the stipulated generations (Section~\ref{eval_hybrid}), whereas the hybrid approaches converged on at least one individual that met the threshold. BEGA Online ran for an average of 17.8 hours, a maximum of 18.32 hours, and a minimum of 17.38 hours before reaching the limit on generations. GAHA Offline took an average of 10.97 hours, a maximum of 11.92 hours and a minimum of 10.31 hours. In contrast, the BEGA Hybrid took an average of 19.1 minutes, a maximum of 24.31 minutes, and a minimum of 16.79 minutes to converge on an optimal solution. The mean absolute difference between the average traffic latency approximated by BENNS and measured on OpenRASE for the qualified individual in BEGA Hybrid was 1ms, underlining BENNS's effectiveness. GENESIS Hybrid took an average of 13.49 minutes, a maximum of 13.78 minutes, and a minimum of 13.25 minutes. GDA took an average of 20.02 s, a maximum of 24.03 s, and a minimum of 18.67 s. The average difference between the average traffic latency estimated by GAHA Offline and the average traffic latency measured on the emulator was 305.72 ms. 

\begin{figure*}
    \centering
    \Description{Line plots showing how GENESIS Hybrid performed in a dynamic network environment.}
    \includegraphics[width=1\linewidth]{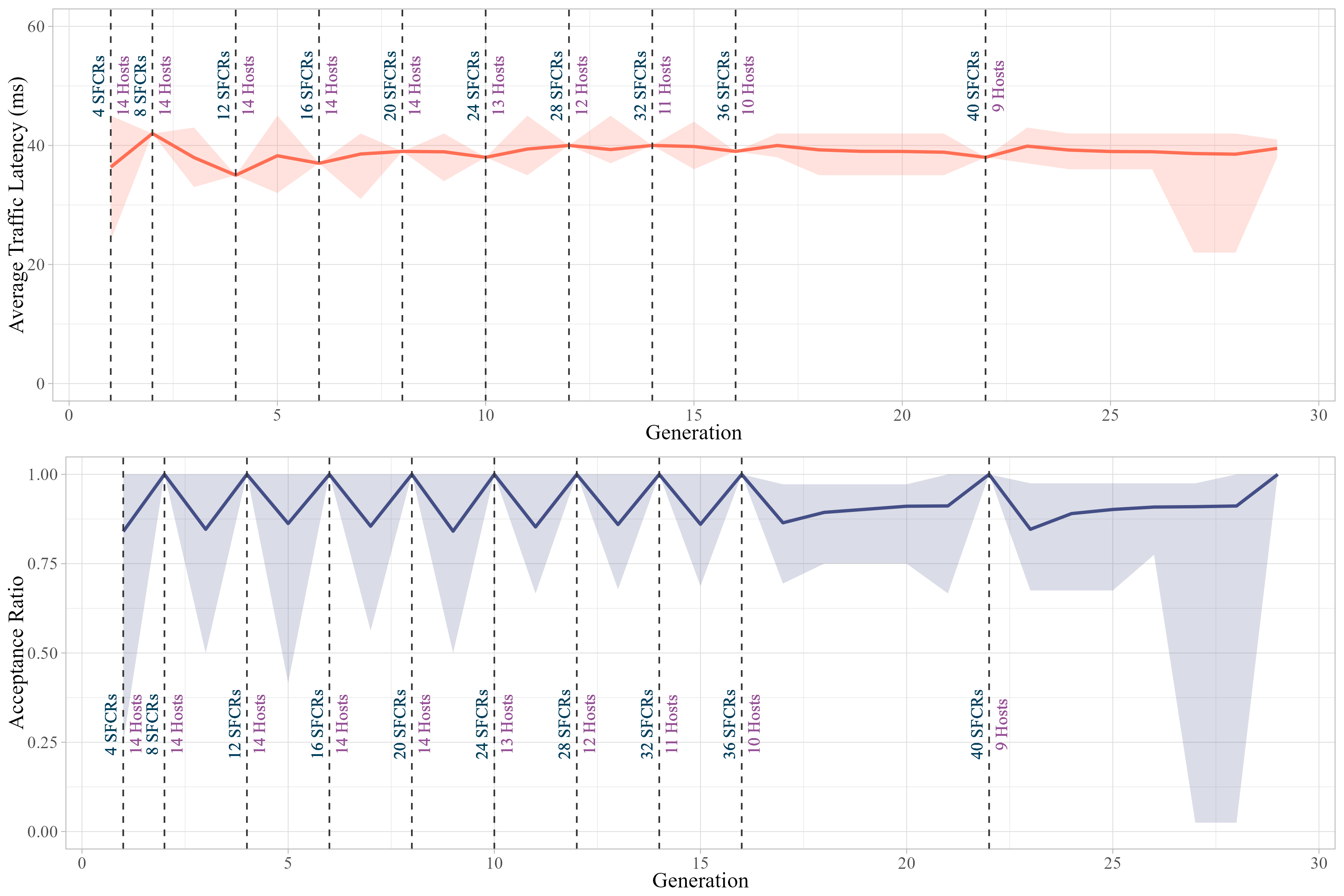}
    \caption{The performance of GENESIS Hybrid in a dynamic network environment. The curves show the average traffic latency and acceptance ratio, while the shaded area shows the maximum and minimum values. The dashed vertical lines segment the evolution based on the state of the network environment. We start with 4 SFCRs and 14 hosts, and by the time of convergence, the state has changed to 8 SFCRs and 14 hosts, and so on. Starting from the 6\textsuperscript{th} segment, hosts crash one by one. The traffic pattern keeps changing every 5 minutes.}
    \label{fig:dynamic}
\end{figure*}

During the evaluation in a dynamic environment, GENESIS Hybrid converged in 8.22 minutes on average following a change in the network environment, such as new SFCRs arriving, traffic pattern changing, or a host crashing. The convergence time increased as the complexity increased, such as when 40 SFCRs had to be embedded with 9 functional hosts, which took 17.62 minutes. Fig.~\ref{fig:dynamic} shows how the average traffic latency and acceptance ratio varied over time during this experiment. 
\section{Discussion}\label{discussion}
\subsection{Empirical Analysis}
The results of the empirical analysis showed that only the maximum CPU and bandwidth demands had an impact on the traffic latency of an SFC. So, we considered only these two values to approximate the traffic latency of SFCs. The maximum memory demand did not impact the traffic latency. Additionally, the negligible increase in maximum memory demand (increased from 1.30 to 1.33), even as the number of requests was ramped up, implies that processing HTTP traffic is not memory-intensive as far as the considered VNFs are concerned. 

\subsection{Evaluation of Hybrid Evolution}
Hybrid online-offline evolution significantly reduced the time taken to produce an optimal solution, allowing evolution with thousands of individuals over hundreds of generations. This enables GAs to explore the search space more, thereby avoiding local optima. While greatly reducing the time taken for fitness evaluation, BENNS also enables parallelisation of fitness evaluation. Consequently, instead of sequentially evaluating the fitness of individuals in a population, the individuals are evaluated in parallel, making evolution over an even greater number of individuals over a greater number of generations possible. The BENNS model was also shown to work across diverse networking scenarios. The traffic latency approximated by BENNS was within 1ms of the traffic latency measured on OpenRASE, demonstrating the model's effectiveness for fitness approximation. Only one online evolution was required before convergence, making online evolution merely a verification step. 

GAHA Offline was slower and less accurate than the hybrid evolutions and did not converge in any experiment because its offline evaluation computed traffic latency mathematically, with no empirical verification. In contrast, BENNS empirically verified its assumptions about factors influencing traffic latency, generated data on an emulator and trained a model to approximate traffic latency, making it more effective. The model, being a simple NN with two hidden layers, enabled BENNS to approximate traffic latency faster than GAHA Offline, whose mathematical model is complicated. 

GENESIS Hybrid was expectedly faster than BEGA Hybrid as it optimised all three sub-problems, providing it with more degrees of freedom, enabling faster exploration of the search space. The use of BENNS with two different GAs (GENESIS Hybrid and BEGA Hybrid) also demonstrated BENNS's ability to be plugged into any GA to enable hybrid evolution. 

The mean acceptance ratios showed a downward trend in some experiments (e.g., Traffic Scale--BEGA Hybrid and Bandwidth--BEGA Hybrid). This is expected because this is a Pareto optimisation problem that tries to meet two opposing objectives. However, the maximum acceptance ratio in BEGA Hybrid kept steadily rising, while the maximum average traffic latency, as shown in Fig.~\ref{fig:traffic_latency}, remained fairly steady after a steep decline in the second generation. It remained steady because the GA, with the aid of BENNS, was able to find SFC embeddings with a high acceptance ratio that did not cause performance bottlenecks. BEGA Online had a lower average traffic latency than the hybrid ones. This is expected because the hybrid experiments had a higher acceptance ratio, which means more SFCs were being embedded. 

The curves showed a sharp change during hybrid evolution for all experiments when transitioning from offline to online evolution. This is because only individuals that met the threshold from offline evolution, only 1 for all experiments, were selected for online evolution, making the curve move sharply. 

The smaller number of individuals in a population in BEGA Online meant that the evolution got stuck in local optima in some experiments (Bandwidth and Traffic Pattern), as can be seen from Fig.~\ref{fig:ar_plot}. In contrast, BENNS allowed hybrid experiments to evolve with 2000 individuals in BEGA Hybrid and 100 individuals in GENESIS Hybrid, allowing them to explore the search space more and eventually converge on an optimal solution. 

The simple and light approximator and the benchmarking, which enable BENNS to approximate traffic latency across diverse environments faster, enabled GENESIS Hybrid to optimise the OSE problem in a dynamic environment effectively, converging on an average of 8.22 minutes following changes in the network environment. BENNS's performance in a dynamic network environment gives it a reasonable chance of being effective in real-life computer networks. 

\section{Conclusion}
This paper introduced a hybrid online-offline approach to evaluate the fitness of GA-based solutions to the NP-hard SFC embedding problem. Unlike other GA-based approaches that use offline (simulation or analytical) or online (emulation) fitness evaluation, our approach combines a novel offline surrogate model called BENNS with an existing online SFC emulator, OpenRASE. Through experimental evaluation across five different network scenarios of varying resources and traffic configurations, and a dynamic network environment, we demonstrate that this approach is both feasible and accurate, exploring tens of thousands of solutions in a matter of minutes to yield an optimal solution to a multi-objective problem, compared to an online-only approach that took several hours to explore tens of solutions but did not converge on an optimal solution, and an offline-only approach which was faster than online but was les accurate.

\printbibliography

@article{Ma2017,
    title = {{A model based on genetic algorithm for service chain resource allocation in NFV}},
    year = {2017},
    journal = {2017 3rd IEEE International Conference on Computer and Communications, ICCC 2017},
    author = {Ma, Ningning and Zhang, Jiao and Huang, Tao},
    month = {7},
    pages = {607--611},
    volume = {2018-Janua},
    publisher = {Institute of Electrical and Electronics Engineers Inc.},
    isbn = {9781509063505},
    doi = {10.1109/COMPCOMM.2017.8322616}
}

@article{Kim2016,
    title = {{An energy-Aware service function chaining {\&} reconfiguration algorithm in NFV}},
    year = {2016},
    journal = {Proceedings - IEEE 1st International Workshops on Foundations and Applications of Self-Systems, FAS-W 2016},
    author = {Kim, Siri and Han, Yunjung and Park, Sungyong},
    month = {12},
    pages = {54--59},
    publisher = {Institute of Electrical and Electronics Engineers Inc.},
    isbn = {9781509036516},
    doi = {10.1109/FAS-W.2016.24}
}

@article{Li2019,
    title = {{An Improved Genetic Algorithm for the Scheduling of Virtual Network Functions}},
    year = {2019},
    journal = {2019 20th Asia-Pacific Network Operations and Management Symposium: Management in a Cyber-Physical World, APNOMS 2019},
    author = {Li, Qi and Wang, Xing and Zhao, Tao and Wang, Ying and Li, Zifan and Rui, Lanlan},
    month = {9},
    publisher = {Institute of Electrical and Electronics Engineers Inc.},
    isbn = {9784885523205},
    doi = {10.23919/APNOMS.2019.8892907}
}

@article{Bunyakitanon2020,
    title = {{Auto-3P: An autonomous VNF performance prediction {\&} placement framework based on machine learning}},
    year = {2020},
    journal = {Computer Networks},
    author = {Bunyakitanon, Monchai and da Silva, Aloizio Pereira and Vasilakos, Xenofon and Nejabati, Reza and Simeonidou, Dimitra},
    month = {11},
    pages = {107433},
    volume = {181},
    publisher = {Elsevier},
    doi = {10.1016/J.COMNET.2020.107433},
    issn = {1389-1286}
}

@article{Fulber-Garcia2024,
    title = {{Breaking the Limits: Bio-Inspired SFC Deployment across Multiple Domains, Clouds and Orchestrators}},
    year = {2024},
    journal = {2024 IEEE Conference on Network Function Virtualization and Software Defined Networks, NFV-SDN 2024},
    author = {Fulber-Garcia, Vinicius and Flauzino, José and Ven{\^{a}}ncio, Giovanni and Huff, Alexandre and Duarte, Elias P.},
    publisher = {Institute of Electrical and Electronics Engineers Inc.},
    isbn = {9798350380538},
    doi = {10.1109/NFV-SDN61811.2024.10807470}
}

@article{Wang2020a,
    title = {{Cost minimization in placing service chains for virtualized network functions}},
    year = {2020},
    journal = {International Journal of Communication Systems},
    author = {Wang, Chien Ting and Lin, Ying Dar and Wang, Chih Chiang and Lai, Yuan Cheng},
    number = {4},
    month = {3},
    pages = {e4222},
    volume = {33},
    publisher = {John Wiley {\&} Sons, Ltd},
    doi = {10.1002/DAC.4222},
    issn = {1099-1131}
}

@article{Fulber-Garcia2023,
    title = {{Customizable Mapping of Virtualized Network Services in Multi-datacenter Environments Based on Genetic Metaheuristics}},
    year = {2023},
    journal = {Journal of Network and Systems Management},
    author = {Fulber-Garcia, Vinicius and Luizelli, Marcelo C. and dos Santos, Carlos R.Paula and Spinosa, Eduardo J. and Duarte, Elias P.},
    number = {4},
    month = {10},
    pages = {1--31},
    volume = {31},
    publisher = {Springer},
    url = {https://link.springer.com/article/10.1007/s10922-023-09760-1},
    doi = {10.1007/S10922-023-09760-1/FIGURES/18},
    issn = {15737705}
}

@article{Qu2016,
    title = {{Delay-Aware Scheduling and Resource Optimization with Network Function Virtualization}},
    year = {2016},
    journal = {IEEE Transactions on Communications},
    author = {Qu, Long and Assi, Chadi and Shaban, Khaled},
    number = {9},
    month = {9},
    pages = {3746--3758},
    volume = {64},
    publisher = {Institute of Electrical and Electronics Engineers Inc.},
    doi = {10.1109/TCOMM.2016.2580150},
    issn = {00906778}
}

@article{Liu2024a,
    title = {{Discrete spider monkey optimization algorithm for latency-sensitive VNF deployment and resource allocation}},
    year = {2024},
    journal = {International Journal of Communication Systems},
    author = {Liu, Xinran and Hou, Yonghong and Liu, Hongchen},
    number = {2},
    month = {1},
    pages = {e5640},
    volume = {37},
    publisher = {John Wiley {\&} Sons, Ltd},
    url = {/doi/pdf/10.1002/dac.5640 https://onlinelibrary.wiley.com/doi/abs/10.1002/dac.5640 https://onlinelibrary.wiley.com/doi/10.1002/dac.5640},
    doi = {10.1002/DAC.5640},
    issn = {1099-1131}
}

@article{Fu2020,
    title = {{Dynamic Service Function Chain Embedding for NFV-Enabled IoT: A Deep Reinforcement Learning Approach}},
    year = {2020},
    journal = {IEEE Transactions on Wireless Communications},
    author = {Fu, Xiaoyuan and Yu, F. Richard and Wang, Jingyu and Qi, Qi and Liao, Jianxin},
    number = {1},
    month = {1},
    pages = {507--519},
    volume = {19},
    publisher = {Institute of Electrical and Electronics Engineers Inc.},
    doi = {10.1109/TWC.2019.2946797},
    issn = {15582248}
}

@article{10255468,
    title = {{Evolutionary Autonomous Networks}},
    year = {2021},
    journal = {Journal of ICT Standardization},
    author = {Harvey, Paul and Tatar, Alexandru and Imai, Pierre and Wong, Leon and Bringuier, Laurent},
    number = {2},
    pages = {201--228},
    volume = {9},
    doi = {10.13052/jicts2245-800X.927}
}

@inproceedings{jin2002fitness,
    title = {{Fitness Approximation In Evolutionary Computation-a Survey.}},
    year = {2002},
    booktitle = {GECCO},
    author = {Jin, Yaochu and Sendhoff, Bernhard},
    pages = {1105--1112},
    volume = {2},
    url = {https://www.honda-ri.de/}
}

@article{Xie2021,
    title = {{FlexChain: Bridging Parallelism and Placement for Service Function Chains}},
    year = {2021},
    journal = {IEEE Transactions on Network and Service Management},
    author = {Xie, Sihao and Ma, Junte and Zhao, Jin},
    number = {1},
    month = {3},
    pages = {195--208},
    volume = {18},
    publisher = {Institute of Electrical and Electronics Engineers Inc.},
    doi = {10.1109/TNSM.2020.3047834},
    issn = {19324537}
}

@article{Ruiz2020,
    title = {{Genetic algorithm for holistic VNF-mapping and virtual topology design}},
    year = {2020},
    journal = {IEEE Access},
    author = {Ruiz, Lidia and Barroso, Ramón J.Durán and De Miguel, Ignacio and Merayo, Noemí and Aguado, Juan Carlos and De La Rosa, Ramón and Fern{\'{a}}ndez, Patricia and Lorenzo, Rubén M. and Abril, Evaristo J.},
    pages = {55893--55904},
    volume = {8},
    publisher = {Institute of Electrical and Electronics Engineers Inc.},
    doi = {10.1109/ACCESS.2020.2982018},
    issn = {21693536}
}

@article{NFV2014,
    title = {{GS NFV 002 - V1.2.1 - Network Functions Virtualisation (NFV); Architectural Framework}},
    year = {2014},
    author = {{NFV}}
}

@inproceedings{Xu2021,
    title = {{HASP: High Availability SFC Placement Approach in Data Center Network}},
    year = {2021},
    booktitle = {2021 IEEE 23rd Int Conf on High Performance Computing {\&} Communications; 7th Int Conf on Data Science {\&} Systems; 19th Int Conf on Smart City; 7th Int Conf on Dependability in Sensor, Cloud {\&} Big Data Systems {\&} Application (HPCC/DSS/SmartCity/DependSys)},
    author = {Xu, Ziqi and Han, Qingmian and Cheng, Bo and Niu, Meng and Chen, Junliang},
    pages = {448--455},
    doi = {10.1109/HPCC-DSS-SmartCity-DependSys53884.2021.00083}
}

@article{6727567,
    title = {{iCloudAccess: Cost-Effective Streaming of Video Games From the Cloud With Low Latency}},
    year = {2014},
    journal = {IEEE Transactions on Circuits and Systems for Video Technology},
    author = {Wu, Di and Xue, Zheng and He, Jian},
    number = {8},
    pages = {1405--1416},
    volume = {24},
    doi = {10.1109/TCSVT.2014.2302543}
}

@article{Jang2017,
    title = {{Joint optimization of service function placement and flow distribution for service function chaining}},
    year = {2017},
    journal = {IEEE Journal on Selected Areas in Communications},
    author = {Jang, Insun and Suh, Dongeun and Pack, Sangheon and D{\'{a}}n, György},
    number = {11},
    month = {11},
    pages = {2532--2541},
    volume = {35},
    publisher = {Institute of Electrical and Electronics Engineers Inc.},
    doi = {10.1109/JSAC.2017.2760162},
    issn = {07338716}
}

@article{Liu2021a,
    title = {{Joint SFC Deployment and Resource Management in Heterogeneous Edge for Latency Minimization}},
    year = {2021},
    journal = {IEEE Transactions on Parallel and Distributed Systems},
    author = {Liu, Yu and Shang, Xiaojun and Yang, Yuanyuan},
    number = {8},
    month = {8},
    pages = {2131--2143},
    volume = {32},
    publisher = {IEEE Computer Society},
    doi = {10.1109/TPDS.2021.3062341},
    issn = {15582183}
}

@article{Harutyunyan2022,
    title = {{Latency and Mobility-Aware Service Function Chain Placement in 5G Networks}},
    year = {2022},
    journal = {IEEE Transactions on Mobile Computing},
    author = {Harutyunyan, Davit and Shahriar, Nashid and Boutaba, Raouf and Riggio, Roberto},
    number = {5},
    month = {5},
    pages = {1697--1709},
    volume = {21},
    publisher = {Institute of Electrical and Electronics Engineers Inc.},
    doi = {10.1109/TMC.2020.3028216},
    issn = {15580660}
}

@article{Gamal2019,
    title = {{Mapping and scheduling for non-uniform arrival of virtual network function (VNF) requests}},
    year = {2019},
    journal = {IEEE Vehicular Technology Conference},
    author = {Gamal, Mahmoud and Jafarizadeh, Saber and Abolhasan, Mehran and Lipman, Justin and Ni, Wei},
    month = {9},
    volume = {2019-Septe},
    publisher = {Institute of Electrical and Electronics Engineers Inc.},
    isbn = {9781728112206},
    doi = {10.1109/VTCFALL.2019.8891197},
    issn = {15502252}
}

@article{Gamal2019a,
    title = {{Mapping and Scheduling of Virtual Network Functions using Multi Objective Optimization Algorithm}},
    year = {2019},
    journal = {Proceedings - 2019 19th International Symposium on Communications and Information Technologies, ISCIT 2019},
    author = {Gamal, Mahmoud and Abolhasan, Mehran and Jafarizadeh, Saber and Lipman, Justin and Ni, Wei},
    month = {9},
    pages = {328--333},
    publisher = {Institute of Electrical and Electronics Engineers Inc.},
    isbn = {9781728150093},
    doi = {10.1109/ISCIT.2019.8905113}
}

@article{SREEKANTH2010245,
    title = {{Multi-objective management of saltwater intrusion in coastal aquifers using genetic programming and modular neural network based surrogate models}},
    year = {2010},
    journal = {Journal of Hydrology},
    author = {Sreekanth, J and Datta, Bithin},
    number = {3},
    pages = {245--256},
    volume = {393},
    doi = {https://doi.org/10.1016/j.jhydrol.2010.08.023},
    issn = {0022-1694}
}

@article{Tavakoli-Someh2019,
    title = {{Multi-objective virtual network function placement using NSGA-II meta-heuristic approach}},
    year = {2019},
    journal = {Journal of Supercomputing},
    author = {Tavakoli-Someh, Sanaz and Rezvani, Mohammad Hossein},
    number = {10},
    month = {10},
    pages = {6451--6487},
    volume = {75},
    publisher = {Springer New York LLC},
    url = {https://dl.acm.org/doi/10.1007/s11227-019-02849-y},
    doi = {10.1007/S11227-019-02849-Y},
    issn = {15730484}
}

@inproceedings{10.1007/978-3-031-33743-7_39,
    title = {{Multi-objective VNF Placement Optimization with NSGA-III}},
    year = {2023},
    booktitle = {Proceedings of the 2023 International Conference on Advances in Computing Research (ACR'23)},
    author = {Bekhit, Mahmoud and Fathalla, Ahmed and Eldesouky, Esraa and Salah, Ahmad},
    editor = {Daimi, Kevin and Al Sadoon, Abeer},
    pages = {481--493},
    publisher = {Springer Nature Switzerland},
    address = {Cham},
    isbn = {978-3-031-33743-7}
}

@article{Toumi2022,
    title = {{On Using Physical Programming for Multi-Domain SFC Placement With Limited Visibility}},
    year = {2022},
    journal = {IEEE Transactions on Cloud Computing},
    author = {Toumi, Nassima and Bernier, Olivier and Meddour, Djamal Eddine and Ksentini, Adlen},
    number = {4},
    month = {10},
    pages = {2787--2803},
    volume = {10},
    publisher = {Institute of Electrical and Electronics Engineers Inc.},
    doi = {10.1109/TCC.2020.3046997},
    issn = {21687161}
}

@inproceedings{OpenRASE,
    title = {{OpenRASE: Service Function Chain Emulation}},
    year = {2025},
    booktitle = {International Conference on Software, Telecommunications and Computer Networks (SoftCOM)},
    author = {Krishnamohan, Theviyanthan and Harvey, Paul},
    url = {https://ieeexplore.ieee.org/document/11197454}
}

@article{Khoshkholghi2019,
    title = {{Optimized Service Chain Placement Using Genetic Algorithm}},
    year = {2019},
    journal = {Proceedings of the 2019 IEEE Conference on Network Softwarization: Unleashing the Power of Network Softwarization, NetSoft 2019},
    author = {Khoshkholghi, Mohammad Ali and Taheri, Javid and Bhamare, Deval and Kassler, Andreas},
    month = {6},
    pages = {472--479},
    publisher = {Institute of Electrical and Electronics Engineers Inc.},
    isbn = {9781538693766},
    doi = {10.1109/NETSOFT.2019.8806644}
}

@article{Rankothge2017,
    title = {{Optimizing Resource Allocation for Virtualized Network Functions in a Cloud Center Using Genetic Algorithms}},
    year = {2017},
    journal = {IEEE Transactions on Network and Service Management},
    author = {Rankothge, Windhya and Le, Franck and Russo, Alessandra and Lobo, Jorge},
    number = {2},
    month = {6},
    pages = {343--356},
    volume = {14},
    publisher = {Institute of Electrical and Electronics Engineers Inc.},
    doi = {10.1109/TNSM.2017.2686979},
    issn = {19324537}
}

@article{Tahmasebi2018,
    title = {{Optimum Transmission Delay for Function Computation in NFV-Based Networks: The Role of Network Coding and Redundant Computing}},
    year = {2018},
    journal = {IEEE Journal on Selected Areas in Communications},
    author = {Tahmasebi, Behrooz and Maddah-Ali, Mohammad Ali and Parsaeefard, Saeedeh and Khalaj, Babak Hossein},
    number = {10},
    month = {10},
    pages = {2233--2245},
    volume = {36},
    publisher = {Institute of Electrical and Electronics Engineers Inc.},
    doi = {10.1109/JSAC.2018.2869952},
    issn = {15580008},
    arxivId = {1807.03337}
}

@article{GilHerrera2016,
    title = {{Resource Allocation in NFV: A Comprehensive Survey}},
    year = {2016},
    journal = {IEEE Transactions on Network and Service Management},
    author = {Gil Herrera, Juliver and Botero, Juan Felipe},
    number = {3},
    month = {9},
    pages = {518--532},
    volume = {13},
    publisher = {Institute of Electrical and Electronics Engineers Inc.},
    doi = {10.1109/TNSM.2016.2598420},
    issn = {19324537}
}

@inproceedings{8647774,
    title = {{Rethinking Fat-Tree Topology Design for Cloud Data Centers}},
    year = {2018},
    booktitle = {2018 IEEE Global Communications Conference (GLOBECOM)},
    author = {Alqahtani, Jarallah and Hamdaoui, Bechir},
    pages = {1--6},
    doi = {10.1109/GLOCOM.2018.8647774}
}

@article{Khoshkholghi2020,
    title = {{Service Function Chain Placement for Joint Cost and Latency Optimization}},
    year = {2020},
    journal = {Mobile Networks and Applications},
    author = {Khoshkholghi, Mohammad Ali and Gokan Khan, Michel and Alizadeh Noghani, Kyoomars and Taheri, Javid and Bhamare, Deval and Kassler, Andreas and Xiang, Zhengzhe and Deng, Shuiguang and Yang, Xiaoxian},
    number = {6},
    month = {12},
    pages = {2191--2205},
    volume = {25},
    publisher = {Springer},
    doi = {10.1007/S11036-020-01661-W/TABLES/5},
    issn = {15728153}
}

@misc{Halpern2015,
    title = {{Service Function Chaining (SFC) Architecture}},
    year = {2015},
    author = {Halpern, Joel M and Pignataro, Carlos},
    number = {7665},
    month = {10},
    series = {Request for Comments},
    publisher = {RFC Editor},
    howpublished = {RFC 7665},
    doi = {10.17487/RFC7665}
}

@article{Pan2019AOptimization,
    title = {{A Classification-Based Surrogate-Assisted Evolutionary Algorithm for Expensive Many-Objective Optimization}},
    year = {2019},
    journal = {IEEE Transactions on Evolutionary Computation},
    author = {Pan, Linqiang and He, Cheng and Tian, Ye and Wang, Handing and Zhang, Xingyi and Jin, Yaochu},
    number = {1},
    month = {2},
    pages = {74--88},
    volume = {23},
    publisher = {Institute of Electrical and Electronics Engineers Inc.},
    doi = {10.1109/TEVC.2018.2802784},
    issn = {1089778X}
}

@article{Sun2019ADesign,
    title = {{A review of the artificial neural network surrogate modeling in aerodynamic design}},
    year = {2019},
    journal = {Proceedings of the Institution of Mechanical Engineers, Part G: Journal of Aerospace Engineering},
    author = {Sun, Gang and Wang, Shuyue},
    number = {16},
    month = {12},
    pages = {5863--5872},
    volume = {233},
    publisher = {SAGE Publications Ltd},
    doi = {https://doi.org/10.1177/0954410019864485},
    issn = {20413025}
}

@article{Shi2010AAlgorithms,
    title = {{A Survey of Fitness Approximation Methods Applied in Evolutionary Algorithms}},
    year = {2010},
    author = {Shi, L. and Rasheed, K.},
    pages = {3--28},
    publisher = {Springer, Berlin, Heidelberg},
    isbn = {978-3-642-10701-6},
    doi = {10.1007/978-3-642-10701-6{\_}1},
    issn = {1867-4542}
}

@misc{Yannan2020AutonomousBusiness,
    title = {{Autonomous Networks, supporting tomorrow's ICT business}},
    year = {2020},
    author = {Yannan, BaiChina and Bingming, Huang and Dong, Sun and John, Strassner and Luigi, Licciardi and Hui, Li and Lei, Wang and Aldo, Artigiani and Dario, Sabella and Haining, Wang and Christian, Maitre and Francisco, Fontes and Yue, Wang and Luca, Pesando and Cecilia, Corbi and Cadzow, Scott},
    url = {https://www.etsi.org/images/files/ETSIWhitePapers/etsi-wp-40-Autonomous-networks.pdf},
    institution = {European Telecommunications Standards Institute}
}

@article{Dushatskiy2019ConvolutionalGOMEA,
    title = {{Convolutional neural network surrogate-assisted GOMEA}},
    year = {2019},
    journal = {GECCO 2019 - Proceedings of the 2019 Genetic and Evolutionary Computation Conference},
    author = {Dushatskiy, Arkadiy and Alderliesten, Tanja and Mendrik, Adriënne M. and Bosman, Peter A.N.},
    month = {7},
    pages = {753--761},
    publisher = {Association for Computing Machinery, Inc},
    isbn = {9781450361118},
    doi = {https://doi.org/10.1145/3321707.3321760}
}

@article{Herker2015Data-centerRequirements,
    title = {{Data-center architecture impacts on virtualized network functions service chain embedding with high availability requirements}},
    year = {2015},
    journal = {2015 IEEE Globecom Workshops, GC Wkshps 2015 - Proceedings},
    author = {Herker, Sandra and An, Xueli and Kiess, Wolfgang and Beker, Sergio and Kirstaedter, Andreas},
    publisher = {Institute of Electrical and Electronics Engineers Inc.},
    isbn = {9781467395267},
    doi = {10.1109/GLOCOMW.2015.7414158}
}

@article{Yang2016Energy-awareCenters,
    title = {{Energy-aware service function placement for service function chaining in data centers}},
    year = {2016},
    journal = {Proceedings - IEEE Global Communications Conference, GLOBECOM},
    author = {Yang, Ke and Zhang, Hong and Hong, Peilin},
    doi = {10.1109/GLOCOM.2016.7841805},
    issn = {25766813}
}

@article{Zu2021FairCenter,
    title = {{Fair Scheduling and Rate Control for Service Function Chain in NFV Enabled Data Center}},
    year = {2021},
    journal = {IEEE Transactions on Network and Service Management},
    author = {Zu, Jiachen and Hu, Guyu and Peng, Dongyang and Xie, Shengxu and Gao, Wenbin},
    number = {3},
    month = {9},
    pages = {2975--2986},
    volume = {18},
    publisher = {Institute of Electrical and Electronics Engineers Inc.},
    doi = {10.1109/TNSM.2021.3070331},
    issn = {19324537}
}

@article{Mori2000GeneticSurvey,
    title = {{Genetic algorithms for adaptation to dynamic environments - A survey}},
    year = {2000},
    journal = {IECON Proceedings (Industrial Electronics Conference)},
    author = {Mori, N. and Kita, H.},
    pages = {2947--2952},
    volume = {1},
    publisher = {IEEE Computer Society},
    doi = {10.1109/IECON.2000.972466}
}

@article{Pandav2022LeveragingCare.,
    title = {{Leveraging 5G technology for robotic surgery and cancer care.}},
    year = {2022},
    journal = {Cancer reports (Hoboken, N.J.)},
    author = {Pandav, Krunal and Te, Austen G and Tomer, Nir and Nair, Sujit S and Tewari, Ashutosh K},
    number = {8},
    month = {8},
    pages = {e1595},
    volume = {5},
    doi = {10.1002/cnr2.1595},
    issn = {2573-8348 (Electronic)},
    pmid = {35266317},
    language = {eng}
}

@article{Oyetunde2022NavigatingSolutions,
    title = {{Navigating kernel selections in kernel-based methods: The issues and possible solutions}},
    year = {2022},
    journal = {AIAA Science and Technology Forum and Exposition, AIAA SciTech Forum 2022},
    author = {Oyetunde, Kehinde S. and Liem, Rhea P.},
    publisher = {American Institute of Aeronautics and Astronautics Inc, AIAA},
    url = {/doi/pdf/10.2514/6.2022-0507},
    isbn = {9781624106316},
    doi = {10.2514/6.2022-0507;ISSUE:ISSUE:10.2514/MSCITECH22;GROUPTOPIC:TOPIC:LONG{\_}TOC{\_}PUBS>LT{\_}ENABLED;CTYPE:STRING:BOOK}
}

@article{Benson2010NetworkWild,
    title = {{Network traffic characteristics of data centers in the wild}},
    year = {2010},
    journal = {Proceedings of the ACM SIGCOMM Internet Measurement Conference, IMC},
    author = {Benson, Theophilus and Akella, Aditya and Maltz, David A.},
    pages = {267--280},
    publisher = {Association for Computing Machinery},
    isbn = {9781450300575},
    doi = {10.1145/1879141.1879175}
}

@article{Ghafariasl2024NeuralSystem,
    title = {{Neural network-based surrogate modeling and optimization of a multigeneration system}},
    year = {2024},
    journal = {Applied Energy},
    author = {Ghafariasl, Parviz and Mahmoudan, Alireza and Mohammadi, Mahmoud and Nazarparvar, Aria and Hoseinzadeh, Siamak and Fathali, Mani and Chang, Shing and Zeinalnezhad, Masoomeh and Garcia, Davide Astiaso},
    month = {6},
    pages = {123130},
    volume = {364},
    publisher = {Elsevier},
    doi = {10.1016/J.APENERGY.2024.123130},
    issn = {0306-2619}
}

@article{Holena2010NeuralAlgorithms,
    title = {{Neural Networks as Surrogate Models for Measurements in Optimization Algorithms}},
    year = {2010},
    journal = {Lecture Notes in Computer Science (including subseries Lecture Notes in Artificial Intelligence and Lecture Notes in Bioinformatics)},
    author = {Holeňa, Martin and Linke, David and Rodemerck, Uwe and Bajer, Lukáš},
    pages = {351--366},
    volume = {6148 LNCS},
    publisher = {Springer, Berlin, Heidelberg},
    isbn = {978-3-642-13568-2},
    doi = {10.1007/978-3-642-13568-2{\_}25},
    issn = {1611-3349}
}

@article{Miller2024OptimizingFidelity,
    title = {{Optimizing composite shell with neural network surrogate models and genetic algorithms: Balancing efficiency and fidelity}},
    year = {2024},
    journal = {Advances in Engineering Software},
    author = {Miller, Bartosz and Ziemia{\'{n}}ski, Leonard},
    month = {11},
    pages = {103740},
    volume = {197},
    publisher = {Elsevier},
    doi = {10.1016/J.ADVENGSOFT.2024.103740},
    issn = {0965-9978}
}

@article{annurev:/content/journals/10.1146/annurev-financial-101620-063859,
    title = {{The Rise of Digital Money}},
    year = {2021},
    journal = {Annual Review of Financial Economics},
    author = {Adrian, Tobias and Mancini-Griffoli, Tommaso},
    number = {Volume 13, 2021},
    pages = {57--77},
    volume = {13},
    publisher = {Annual Reviews},
    url = {https://www.annualreviews.org/content/journals/10.1146/annurev-financial-101620-063859},
    doi = {https://doi.org/10.1146/annurev-financial-101620-063859},
    issn = {1941-1375}
}

@article{Yuan2018,
    title = {{Virtual network function scheduling via multilayer encoding genetic algorithm with distributed bandwidth allocation}},
    year = {2018},
    journal = {Science China Information Sciences},
    author = {Yuan, Quan and Tang, Hongbo and You, Wei and Wang, Xiaolei and Zhao, Yu},
    number = {9},
    month = {9},
    pages = {1--19},
    volume = {61},
    publisher = {Science in China Press},
    url = {https://link.springer.com/article/10.1007/s11432-017-9357-7},
    doi = {10.1007/S11432-017-9357-7/METRICS},
    issn = {18691919}
}

@article{Carpio2017,
    title = {{VNF placement with replication for Load balancing in NFV networks}},
    year = {2017},
    journal = {IEEE International Conference on Communications},
    author = {Carpio, Francisco and Dhahri, Samia and Jukan, Admela},
    month = {7},
    publisher = {Institute of Electrical and Electronics Engineers Inc.},
    isbn = {9781467389990},
    doi = {10.1109/ICC.2017.7996515},
    issn = {15503607},
    arxivId = {1610.08266}
}

@article{Cao2017,
    title = {{VNF-FG design and VNF placement for 5G mobile networks}},
    year = {2017},
    journal = {Science China Information Sciences},
    author = {Cao, Jiuyue and Zhang, Yan and An, Wei and Chen, Xin and Sun, Jiyan and Han, Yanni},
    number = {4},
    month = {4},
    pages = {1--15},
    volume = {60},
    publisher = {Science in China Press},
    doi = {10.1007/S11432-016-9031-X/METRICS},
    issn = {1674733X}
}

@incollection{Fahmy2023SimulatorsWSNs,
    title = {{Simulators and Emulators for WSNs}},
    year = {2023},
    booktitle = {Concepts, Applications, Experimentation and Analysis of Wireless Sensor Networks},
    author = {Fahmy, Hossam Mahmoud Ahmad},
    pages = {547--663},
    publisher = {Springer Nature Switzerland},
    address = {Cham},
    isbn = {978-3-031-20709-9},
    doi = {10.1007/978-3-031-20709-9{\_}8}
}

@article{2013Software-DefinedIt,
    title = {{Software-Defined Networking: Why We Like It and How We Are Building On It}},
    year = {2013},
    url = {https://www.cisco.com/c/dam/en_us/solutions/industries/docs/gov/cis13090_sdn_sled_white_paper.pdf}
}

@article{Acampora2025SolvingAlgorithms,
    title = {{Solving the Ising Problem by Noisy Quantum Genetic Algorithms}},
    year = {2025},
    author = {Acampora, Giovanni and Minolfi, Giulio and Schiattarella, Roberto},
    month = {6},
    pages = {1--7},
    publisher = {Institute of Electrical and Electronics Engineers (IEEE)},
    doi = {10.1109/MCII64973.2025.11032994}
}

@techreport{Imai2020TowardsNetwork,
    title = {{Towards A Truly Autonomous Network}},
    year = {2020},
    author = {Imai, Pierre and Harvey, Paul and Amin, Tareq},
    institution = {Rakuten Mobile Innovation Studio}
}

\end{document}